\documentclass[acmsmall,screen,nonacm]{acmart}

\newcommand{\purple}[1]{#1}

\DeclareRobustCommand{\LinearW}{\purple{Linear}}
\newcommand{\linear}{\purple{linear}}
\DeclareRobustCommand{\BranchingW}{\purple{Branching}}
\newcommand{\branching}{\purple{branching}}
\newcommand{\Commit}[1]{\ensuremath{\mathcal{C}_{#1}}}

\usepackage{listings}
\usepackage{color}
\usepackage{siunitx}
\usepackage{enumitem}
\usepackage{paralist}

\definecolor{JMMSyncAction}{rgb}{0.3,0.7,1}
\definecolor{communicationblue}{rgb}{0.2,0.4,1}

\newcommand{\TightenPar}[1]{\looseness=-#1}

\usepackage{mathpartir}
\usepackage{microtype}
\usepackage[scale=0.9, txmd]{plex-mono}
\usepackage{subcaption}
\usepackage{multirow}

\usepackage[capitalise,nameinlink]{cleveref}
\crefname{section}{§$\!$}{§§$\!$}

\usepackage{etoolbox}
\AtBeginEnvironment{lemma}{\crefalias{theorem}{lemma}}
\AtBeginEnvironment{definition}{\crefalias{theorem}{definition}}

\usepackage{todonotes}
\usepackage{wrapfig}
\usepackage{pifont}
\usepackage{csquotes}

\newcommand{\commentcolor}{magenta}

\definecolor{lukeorange}{rgb}{0.85,0.45,0.0}

\usepackage{colortbl}
\usepackage{multicol}
\usepackage{refcount}
\renewcommand{\c}[1]{{\footnotesize\texttt{#1}}}
\lstnewenvironment{code}[1][]{\lstset{language=Java,numbers=left,#1}}{}
\lstnewenvironment{smallcode}[1][]{\lstset{language=Java,style=SmallCode,numbers=left,#1}}{}

  \newlength{\syncarrowwidth}
  \newcommand{\Pull}[2]{\ensuremath{#1 \mathrel{\makebox[\syncarrowwidth][l]{$\longleftarrow$}} #2}}
  \newcommand{\Push}[2]{\ensuremath{#1 \mathrel{\makebox[\syncarrowwidth][l]{$\longrightarrow$}} #2}}
  \newcommand{\Pullpush}[2]{
    \ensuremath{#1%
      \mathrel{%
        \makebox[\syncarrowwidth][l]{%
          \tikz[baseline=-0.5ex, x=\syncarrowwidth]{
            \draw[{Classical TikZ Rightarrow}-,bend left=40, line width=0.7pt] (0,0) to (1,0.02);
            \draw[{Classical TikZ Rightarrow}-,bend left=40, line width=0.7pt] (1,0) to (0,-0.02);
          }
        }
      }%
    #2}
  }

\lstnewenvironment{threadcode}[1][]{%
  \lstset{language=Java, morekeywords={fork,join,lock,unlock},
          basicstyle=\fontsize{7}{8.4}\selectfont\tt\color{black},
          keywordstyle=\fontsize{7}{8.4}\selectfont\tt\bfseries,
          numbers=none, autogobble, xleftmargin=0.4ex, aboveskip=0pt, belowskip=0pt, #1}}{}

\newlength{\threadcolwidth}
\newenvironment{threadtable}%
  {\begin{tabular}[t]{@{}l@{\hspace{0.5em}}|@{\hspace{0.5em}}l@{}}}%
  {\end{tabular}}
\newcommand{\threadhead}{%
  \hfil\footnotesize Thread 1\hfil &
  \hfil\footnotesize Thread 2\hfil\\
  \midrule}

\newcommand\hlbox[2]{%
  {\setlength\fboxsep{1pt}\colorbox{#1}{\mathstrut #2}}}

\colorlet{smhl}{yellow!50}
\colorlet{sashl}{teal!25}
\newcommand{\sm}[1]{\hlbox{smhl}{#1}}
\newcommand{\sas}[1]{\hlbox{sashl}{#1}}

\newlength{\verdictwidth}
\newlength{\verdictrowheight}
\newlength{\threadcellheight}
\newenvironment{threadcell}%
  {\ifdim\threadcellheight>0pt
     \minipage[t][\threadcellheight][t]{\threadcolwidth}%
   \else
     \minipage[t]{\threadcolwidth}%
   \fi
   \vspace{-0.75em}}%
  {\endminipage}
\newlength{\threadpanelheight}
\newenvironment{threadpanel}%
  {\begin{minipage}[t][\threadpanelheight][t]{\linewidth}\centering}%
  {\end{minipage}}

\usepackage{lstautogobble}

\lstdefinestyle{SmallCode}{
  basicstyle=\fontsize{7}{9}\selectfont\tt\color{black},
  keywordstyle=\fontsize{7}{9}\selectfont\tt\bfseries,
  commentstyle=\color{\commentcolor}\it,
  numberstyle=\fontsize{5}{10}\selectfont\tt\color{black},
  frame=none,
  framerule=0.5pt,
  xleftmargin=5ex,
  xrightmargin=\fboxsep,
  escapeinside={/*Latex:}{*/},
  showstringspaces=false,
  numbers=left,
  columns=fullflexible,
  breakatwhitespace=false,
  breaklines=false,
  belowskip=1ex,
  aboveskip=1ex
}

\usepackage{mathtools}

\newcommand{\Conf}[3]{\langle #1 \,;\, #2 \,;\, #3 \rangle} 
\newcommand{\upd}[3][]{[\,#2 \mathrel{\mapsto\if\relax\detokenize{#1}\relax\else_{#1}\fi} #3\,]}
\newcommand{\anc}[3][]{[\,#2 \mathrel{\leftarrow\if\relax\detokenize{#1}\relax\else_{#1}\fi} #3\,]}
\newcommand{\dom}{\mathsf{dom}}
\newcommand{\holds}{\mathsf{held}}
\renewcommand{\unit}{\epsilon}
\newcommand{\step}{\longrightarrow}       
\newcommand{\emit}[1]{\xRightarrow{#1}}   
\newcommand{\apply}[2]{#1\langle #2\rangle} 
\newcommand{\Val}{\mathsf{Val}}
\newcommand{\Exc}{\mathsf{Exc}}
\newcommand{\Exp}{\mathsf{Exp}}         
\newcommand{\TID}{\mathsf{TID}}
\newcommand{\Var}{\mathsf{Var}}
\newcommand{\Lock}{\mathsf{Lock}}
\newcommand{\CommitID}{\mathsf{CommitID}}
\newcommand{\kw}[1]{\mathsf{#1}}
\newcommand{\push}[1]{\xrightarrow{\text{push}_{#1}}} 
\newcommand{\pull}[1]{\xrightarrow{\text{pull}_{#1}}} 
\newcommand{\commit}[1]{\xrightarrow{\text{commit}_{#1}}} 
\newcommand{\merge}[2]{\xrightarrow{\text{merge}_{#1, #2}}} 
\newcommand{\changes}[2]{\Delta_{#1}^{#2}} 

\newcommand{\new}{\mathcal{N}}       
\newcommand{\running}{\mathcal{R}} 
\newcommand{\terminated}{\mathcal{T}} 

\newcommand{\Locks}{L}

\newcommand{\Heap}{H}

\newcommand{\Versioned}[1]{\colorbox{violet!25}{\ensuremath{#1}}}

\newcommand{\Read}[0]{\ensuremath{\mathsf{read}}}
\newcommand{\Init}[0]{\ensuremath{\mathsf{init}}}
\newcommand{\Write}[0]{\ensuremath{\mathsf{write}}}
\newcommand{\Start}[0]{\ensuremath{\mathsf{start}}}
\newcommand{\End}[0]{\ensuremath{\mathsf{end}}}
\newcommand{\Spawn}[0]{\ensuremath{\mathsf{spawn}}}
\newcommand{\EJoin}[0]{\ensuremath{\mathsf{join}}}
\newcommand{\ELock}[0]{\ensuremath{\mathsf{lock}}}
\newcommand{\Unlock}[0]{\ensuremath{\mathsf{unlock}}}

\newcommand{\Log}[0]{\ensuremath{\mathit{log}}}
\newcommand{\Play}[0]{\ensuremath{\rhd}}
\newcommand{\View}[0]{\ensuremath{\mathit{view}}}

\usepackage{array}
\newcolumntype{L}{>{\small\raggedright\arraybackslash\hyphenpenalty=10000\exhyphenpenalty=10000}p{3.0cm}}
\newcolumntype{C}{>{\small\centering\arraybackslash}p{3.2cm}}
\newcolumntype{A}{>{\fontsize{8}{10}\selectfont\raggedright\arraybackslash\hyphenpenalty=10000\exhyphenpenalty=10000}p{1.65cm}}
\newcolumntype{B}{>{\fontsize{8}{10}\selectfont\raggedright\arraybackslash}p{2.64cm}}
\newcolumntype{E}{>{\fontsize{8}{10}\selectfont\centering\arraybackslash\hyphenpenalty=10000\exhyphenpenalty=10000}p{1.2cm}}
\newcolumntype{F}{>{\fontsize{8}{10}\selectfont\centering\arraybackslash}p{2.45cm}}
\newcolumntype{Z}{>{\fontsize{8}{10}\selectfont\raggedright\arraybackslash}p{3.25cm}}

\newcommand{\nb}[3]{
    {\colorbox{#2}{\bfseries\sffamily\scriptsize\textcolor{white}{#1}}}{
        \textcolor{#2}{\sf\small\textit{#3}}}}

\def\showtodos{}
\ifdefined\showtodos
\newcommand{\DBS}[1]{\todo[inline, size=\small,noline,color=blue!30]{\textbf{DBS:} #1}}
\newcommand{\TW}[1]{\todo[inline, size=\small,noline,color=magenta!30]{\textbf{TW:} #1}}
\newcommand{\FS}[1]{\todo[inline, size=\small,noline,color=orange!30]{\textbf{FS:} #1}}
\newcommand{\tw}[1]{\nb{TW}{magenta}{#1}}
\newcommand{\fs}[1]{\nb{FS}{orange}{#1}}
\newcommand{\lc}[1]{\nb{LC}{orange}{#1}}
\newcommand{\dbs}[1]{\nb{DBS}{blue!60}{#1}}

\newcommand{\EC}[1]{\todo[inline, size=\small,noline,color=green!60]{\textbf{EC:} #1}}
\newcommand{\ec}[1]{\nb{EC}{green!60}{#1}}
\else
\newcommand{\DBS}[1]{}
\newcommand{\TW}[1]{}
\newcommand{\FS}[1]{}
\newcommand{\tw}[1]{}
\newcommand{\fs}[1]{}
\newcommand{\lc}[1]{}
\newcommand{\dbs}[1]{}
\newcommand{\EC}[1]{}
\newcommand{\ec}[1]{}
\fi

\newcommand{\ie}{\emph{i.e.,}}

\newcommand{\cf}{\emph{cf.,}}

\newcommand{\SyncasShare}{Sync-as-Share}

\usetikzlibrary{automata,shapes,decorations,arrows,calc,arrows.meta,fit,positioning,quotes,tikzmark,shadows}

\newlength{\codeboxleftshift}
\newlength{\codeboxrightshift}
\tikzset{
  common timeline styles/.style={
    >={Latex[scale=0.8]},
    every node/.style={font=\scriptsize\ttfamily},
    codebox/.style={
      rounded corners=1.5pt,
      inner sep=1pt,
    },
    thread lane/.style={ultra thick, draw=threadtime, dashed},
    shared lane/.style={thread lane, draw=sharedmem},
    history link/.style={line width=0.8pt},
    link left/.style={history link, {Latex[round, scale=1.1]}-, draw=communicationblue},
    link right/.style={history link, -{Latex[round, scale=1.1]}, draw=communicationblue},
    link oneway/.style={history link, -{Latex[round, scale=1.1]}, draw=communicationblue},
    link both/.style={history link, {Latex[round, scale=1.1]}-{Latex[round, scale=1.1]}, draw=communicationblue},
    conflict/.style={line width=3pt, -{Latex[round,scale=0.8]}, draw=conflictred},
    stateupdate/.style={
      draw=threadtime,
      fill=threadtime!15,
      rounded corners=4pt,
      inner sep=1.5pt,
      font=\scriptsize\ttfamily
    },
    sharedupdate/.style={
      draw=sharedmem!80!black,
      fill=sharedmem!30,
      rounded corners=4pt,
      inner sep=1.5pt,
      font=\scriptsize\ttfamily
    },
    msg/.style={stateupdate, minimum width=13pt, minimum height=9pt},
    msg right/.style={msg, right, xshift=3pt},
    msg left/.style={msg, left, xshift=-3pt},
    message/.style={msg},
    open link/.style={link oneway, dashed},
    question/.style={font=\small\bfseries\sffamily, text=communicationblue, inner sep=1.5pt},
    t1 code/.style={
      anchor=west,
      xshift=\the\codeboxleftshift,
      codebox
    },
    t2 code/.style={
      anchor=west,
      xshift=\the\codeboxrightshift,
      codebox
    },
    t1 state/.style={right, xshift=3pt,  stateupdate},
    t2 state/.style={left,  xshift=-3pt, stateupdate},
    g state right/.style={right, xshift=3pt, sharedupdate},
    g state left/.style={left,  xshift=-3pt, sharedupdate},
  }
}

\newcommand{\TimelineColorSetup}{
  \definecolor{threadtime}{rgb}{0.5,0.5,0.5}
  \definecolor{sharedmem}{rgb}{0.3,0.3,0.3}
  \definecolor{conflictred}{rgb}{1,0.2,0.2}
}

\newcommand{\DefineThreadEventMacros}{
  \newcommand{\ResolveThreadX}[1]{\def\tx{\onex}\ifnum\pdfstrcmp{##1}{t2}=0\def\tx{\twox}\fi}
  \newcommand{\ThreadEvent}[4]{\ResolveThreadX{##1}\fill[fill=threadtime] (\tx,##2) circle (2pt) coordinate (##3) node[##1 code] {##4};}
  \newcommand{\ThreadEventFrom}[5]{\fill[fill=threadtime] ($(##2)+(0,##3)$) circle (2pt) coordinate (##4) node[##1 code] {##5};}
  \newcommand{\ThreadSyncEvent}[4]{\ResolveThreadX{##1}\fill (\tx,##2) circle (0pt) coordinate (##3) node[##1 code] {##4};}
  \newcommand{\ThreadSyncEventFrom}[5]{\fill ($(##2)+(0,##3)$) circle (0pt) coordinate (##4) node[##1 code] {##5};}
  \newcommand{\PullPush}[2]{
    \path let \p1 = ##1, \p2 = ##2 in \pgfextra{%
      \ifdim\x1<\x2
        \draw[link oneway] ##2 to[looseness=.5, out=140, in=30] ##1;
        \draw[link oneway] ##1 to[looseness=.5, out=320, in=210] ##2;
      \else
        \draw[link oneway] ##2 to[looseness=.5, out=40, in=150] ##1;
        \draw[link oneway] ##1 to[looseness=.5, out=220, in=330] ##2;
      \fi
    };
  }
  \newcommand{\PullPushMsg}[3]{
    \path let \p1 = ##1, \p2 = ##2 in \pgfextra{%
      \ifdim\x1<\x2
        \draw[link oneway] ##2 to[looseness=.5, out=140, in=30] ##1;
        \draw[link oneway] ##1 to[looseness=.5, out=320, in=210] node[message] {##3} ##2;
      \else
        \draw[link oneway] ##2 to[looseness=.5, out=40, in=150] ##1;
        \draw[link oneway] ##1 to[looseness=.5, out=220, in=330] node[message] {##3} ##2;
      \fi
    };
  }
  \newcommand{\ConflictEvent}[2]{
    \node[regular polygon, regular polygon sides=8,
        draw=black, fill=black, line width=2pt,
        minimum size=20pt, inner sep=0pt] (##2) at (##1) {};
    \node[regular polygon, regular polygon sides=8,
        draw=white, fill=conflictred, line width=0.9pt,
        minimum size=14pt, inner sep=0pt,
        font=\scriptsize\bfseries\sffamily, text=white] at (##1) {\textsc{fail}};
  }
}

\newcommand{\CodeWidthOf}[1]{\widthof{{\scriptsize\ttfamily\selectfont #1}}}
\newcommand{\SetLeftCodeShift}[1]{%
  \settowidth{\codeboxleftshift}{#1}%
  \setlength{\codeboxleftshift}{-\codeboxleftshift}%
  \addtolength{\codeboxleftshift}{-3mm}%
}
\newcommand{\SetRightCodeShift}[1]{%
  \setlength{\codeboxrightshift}{#1}%
}

\makeatletter
\lst@AddToHook{Init}{\def\@currentcounter{lstnumber}}%
\makeatother

\begin{document}

\title{Nothing Is Shared Until You Synchronise}
\subtitle{Reinterpreting Java for Data-Race Freedom}
\setcounter{page}{1}
\author{Luke Cheeseman}
\orcid{0009-0000-0632-9825}
\affiliation{%
  \institution{Uppsala University}
  \city{Uppsala}
  \country{Sweden}
}
\email{luke.cheeseman@it.uu.se}

\author{Elias Castegren}
\orcid{0000-0003-4918-6582}
\affiliation{%
  \institution{Uppsala University}
  \city{Uppsala}
  \country{Sweden}
}
\email{elias.castegren@it.uu.se}

\author{David Black-Schaffer}
\orcid{0000-0001-5375-4058}
\affiliation{%
  \institution{Uppsala University}
  \city{Uppsala}
  \country{Sweden}
}
\email{david.black-schaffer@it.uu.se}

\author{Erik Österlund}
\orcid{0000-0003-3686-8568}
\affiliation{%
  \institution{Oracle}
  \city{Stockholm}
  \country{Sweden}
}
\email{erik.osterlund@oracle.com}

\author{Tobias Wrigstad}
\orcid{0000-0002-4269-5408}
\affiliation{%
  \institution{Uppsala University}
  \city{Uppsala}
  \country{Sweden}
}
\email{tobias.wrigstad@it.uu.se}

\begin{abstract}
In mainstream shared-memory languages, programmers use synchronisation mechanisms such as locks to isolate shared data accesses across threads to avoid data races.
Yet synchronisation enforces only a temporal ordering, while the underlying shared memory continues to communicate updates between threads at every ordinary read and write.
To achieve isolation, programmers therefore need to understand the connection between ordering and communication, and to apply synchronisation correctly and consistently on that basis, while compilers and hardware optimise assuming they have.
When programmers do not order (synchronise) correctly, they lose isolation, and the semantics of their programs change without notice.

We propose \SyncasShare{}, a memory model built on isolation and communication rather than on the ordering of accesses to shared memory.
We start from thread isolation, where each thread reads and writes a private view of memory, and reinterpret programs' existing synchronisation actions to take on the additional role of explicitly communicating updates.
As nothing is communicated outside of synchronisation actions, data races become impossible by construction.
Instead, conflicting updates are now detected at synchronisation points.
Importantly, by viewing synchronisation as explicit communication, we are able to identify, expose, and address the underlying synchronisation design decisions of what should be communicated, to whom, and what counts as a conflict.
We explore these design decisions in \SyncasShare{} and propose two different communication topologies with different trade-offs.
To make the implications concrete, we instantiate \SyncasShare{} for Java by making the synchronisation actions already defined by its memory model the only points at which threads communicate.
Java programs can therefore adopt \SyncasShare{} unmodified.
We then formalise this instantiation, drawing on version control systems to build version-controlled memory, in which synchronisation actions push and pull updates between private views, and mechanise the formalisation in Rocq.
From this we can then show that Java programs that are correctly synchronised (as defined by the current Java memory model) behave identically under shared memory and \SyncasShare{}, while providing the benefits of isolation and explicit communication.

\end{abstract}

\maketitle

\section{Introduction}
\label{sec:introduction}

Programmers using shared memory aim to work in \emph{isolation} and to reason about a thread as if the state it touches were its own.
They use synchronisation at the boundaries where that isolation must be broken, so that data can pass from one thread to another.
Locks, volatiles, and thread start/join are the vocabulary for those boundaries. But they do not deliver isolation. Instead they enforce a \emph{temporal~ordering} which is only indirectly connected to isolation. Every thread still reads and writes one shared memory, and the mechanisms order threads without denoting the state being handed over.
This misalignment places two demands on programmers:
they must apply synchronisation correctly and consistently across all accesses to all shared data, and they must understand the language and hardware memory models well enough to know how ordering in time translates into isolation across space.
This has proven difficult: programmers do not consistently apply synchronisation correctly~\cite{LearningFromMistakes, BoehmBenign}, the races that result are hard enough to find that a whole class of tools exists to find them~\cite{eraser, FastTrack, 10.1145/3759548.3763369}, and understanding how ordering translates into isolation for incorrectly synchronised programs is itself a research topic~\cite{JMM, SevcikAspinall2008, Batty2011}.
Failing either demand has a cost.
Compilers and hardware assume synchronisation is correct, and the optimisations they apply preserve semantics only when that assumption holds~\cite{JMM, BoehmAdve2008}.
To avoid these problems entirely, some languages and concurrency paradigms enforce isolation by construction, but they require programs to be (re-)written following their semantics, which is prohibitively expensive for large existing projects based on shared memory.

In this paper, we explore the consequences of fixing the misalignment of synchronisation and isolation for programs written against shared memory.
We do so by starting from \emph{thread~isolation} rather than globally shared memory.
In our approach, each thread operates on a private view of memory, and synchronisation operations (such as unlock) are reinterpreted to provide both explicit ordering and \emph{explicit~communication} of shared data.
By explicit communication, we mean that a thread's synchronisation points are the only places where its private updates are communicated outwards and other threads' updates inwards.

This memory model has several important properties.
First, isolation means that data races in the shared-memory sense are impossible by construction.
Explicit communication means that a program's synchronisation dictates precisely where updates from different threads are brought together, and any \emph{conflicts} (such as unsynchronised concurrent writes to the same variables) are detected there.
Second, explicit communication exposes the design decisions of what data is shared with which threads, what we call the communication \emph{topology}, and what constitutes a conflict.
Finally, extending existing program synchronisation ordering to provide explicit communication means that programs need no changes and correctly synchronised programs retain their semantics.

We call this memory model \SyncasShare{}.
For concreteness, we instantiate \SyncasShare{} for Java by reinterpreting the synchronisation actions of the Java Memory Model (JMM), \ie{} monitors, volatile fields, and thread start and join.
In this paper:
\begin{compactitem}
\item we introduce \SyncasShare{} and explore the consequences of defaulting to thread isolation with synchronisation as explicit communication (\cref{sec:syncasshare-intro});
\item we explore realisations for two communication topologies (\cref{sec:syncasshare});
\item we formalise the semantics of the two topologies together with a simplified shared-memory baseline (\cref{sec:vcm});
\item we mechanise our formalism in Rocq~\cite{vcmRocq} and show that correctly synchronised programs under \SyncasShare{} present the same behaviour as under shared memory (\cref{sec:metatheory});
\item we discuss some of the design choices inherent to \SyncasShare{} (\cref{sec:design}).
\end{compactitem}

Our aim in this paper is to design a memory model built on isolation and communication, rather than on the ordering of accesses to shared memory.
The motivation is to enable a better understanding of, and avoidance of, data races, while providing backwards compatibility for correctly synchronised programs through the reuse of their existing synchronisation operations.
In particular, this paper does not address implementation and optimisation of the model, which we leave for future work.
While we understand that the most direct implementation of \SyncasShare{} would have significant overhead, we encourage the reader to consider this akin to the first introduction of other quality-of-code ideas, such as garbage collection, whose overheads have been successfully addressed over time.
In addition, while our design is largely language agnostic, we explore our model in the context of Java, and explicitly leave discussions of how it would interact with other languages' memory models for later work.

\section{Sync-as-Share}
\label{sec:syncasshare-intro}

\SyncasShare{} has two fundamental properties: \emph{thread isolation} and \emph{explicit communication}.
Beyond those, there are a range of design decisions that any practical implementation needs to address: 1) which \emph{actions} communicate, 2) \emph{what} they communicate, 3) with \emph{whom} they communicate, 4) which \emph{direction} information flows, and, 5) how a thread \emph{reconciles} communication.
These design decisions are not limited to \SyncasShare{}.
Indeed, this list provides a general roadmap for exploring the interaction between concurrency and communication, as can be seen in \cref{tab:communications-models}, which lists the implicit choices that popular programming models have made in each of these dimensions.

\begin{table}[h]
\caption{Communication choices in popular programming models and \SyncasShare{}.}
\label{tab:communications-models}
\centering
\setlength{\tabcolsep}{3pt}
\begin{tabular}{A>{\columncolor{smhl}}BBB>{\columncolor{sashl}}Z}
\toprule
 & \textbf{Shared Memory}                              & \textbf{Transactions (STM)}            & \textbf{Actors}                      & \textbf{\SyncasShare{}}                            \\
\midrule
(1) actions    & All load and store instructions                          & Transactional variable reads and commits  & Message sends and receives           & Explicit program synchronisation actions                                  \\
\arrayrulecolor{black!10}\midrule\arrayrulecolor{black}

(2) what       & All changes to the cacheline                             & The values of trans- actional variables     & Message contents                     & Thread's whole view                                      \\
\arrayrulecolor{black!10}\midrule\arrayrulecolor{black}
(3) whom       & All other threads                                        & All other threads                         & The target actor                     & All threads, or users of the same synchronisation object \\
\arrayrulecolor{black!10}\midrule\arrayrulecolor{black}
(4) direction  & Load (in)\par Store (out)                                & Read (in)\par Commit (out)                & Receive (in)\par Send (out)          & Acquire (in)\par Release (out, or both)                  \\
\arrayrulecolor{black!10}\midrule\arrayrulecolor{black}
(5) reconciled & Writes overwrite; reads see what the memory model allows & Abort on write to a read or written location & Objects in message's payload added to actor's heap & Conflict on writes to the same location                  \\
\bottomrule
\end{tabular}
\end{table}


When we instantiate \SyncasShare{} for Java, we must adhere to the guarantees given by the JMM: all executions of a correctly synchronised program must appear sequentially consistent~\cite[\S17.4.5]{JMM}.
For correctly synchronised programs we explicitly aim to preserve all valid executions (\cf{} \cref{thm:equiv}).
That requirement answers the first two design decisions outright, sets a minimum on the next two, and leaves the last to us.
It also bounds our design choices, as they may only affect how an incorrectly synchronised program behaves, and where its conflicts surface.
Our mapping onto Java therefore works thus:
1)~The actions that communicate are Java's synchronisation actions~\cite[\S17.4.2]{JMM}: locking and unlocking a monitor, reading and writing a volatile field, starting and joining a thread, and the synthetic first and last actions of each thread.\footnote{We do not include VarHandles' access modes~\cite{varhandlesjep}.}
2)~The thread's whole view is communicated, since anything narrower would break guarantees Java makes today~\cite[\S17.4.5]{JMM}.
3)~Java requires that two threads synchronising on the same object communicate~\cite[\S17.4.4]{JMM}, but a thread that communicates more widely still meets that requirement.
How narrowly or widely threads communicate is the \emph{communication topology}, and we present two options in \cref{sec:syncasshare}.
4)~Java's synchronisation actions either release or acquire~\cite[\S17.4.4]{JMM}: a release communicates outwards, an acquire inwards.
\SyncasShare{} must match the communication directions these actions provide, but may additionally communicate in the other direction.
5)~Deciding what counts as a conflict is a choice that is entirely ours, as Java does not detect nor prohibit data races.
The last point is where we choose to break backwards-compatibility with the JMM's ignoring conflicts: we consider updates to the same location in two views to be a conflict, regardless of their values, and throw an exception when such a conflict is detected.
We revisit alternative choices for this design space, including program-defined conflict definitions, in \cref{sec:design}.

\begin{figure}[t]
  \centering
  \captionsetup[subfigure]{position=top}
  {\footnotesize \textbf{Note:} at the start of all threads \c{x=y=a=b=0}}
  \par\smallskip
  \begin{subfigure}[t]{0.32\textwidth}
    \centering
    \caption{Write isolation}
    \label{fig:fork-no-join}
    \begin{threadpanel}
    \begin{threadtable}
      \threadhead
      \begin{threadcell}
        \begin{threadcode}
          x = 1
        \end{threadcode}
      \end{threadcell}
      &
      \begin{threadcell}
        \begin{threadcode}
          x = 2
        \end{threadcode}
      \end{threadcell}
    \end{threadtable}
    \end{threadpanel}
  \end{subfigure}
  \hfill
  \begin{subfigure}[t]{0.32\textwidth}
    \centering
    \caption{Write conflict}
    \label{fig:fork-join-write-write}
    \begin{threadpanel}
    \begin{threadtable}
      \threadhead
      \begin{threadcell}
        \begin{threadcode}
          x = 1
          join T2
        \end{threadcode}
      \end{threadcell}
      &
      \begin{threadcell}
        \begin{threadcode}
          x = 1
        \end{threadcode}
      \end{threadcell}
    \end{threadtable}
    \end{threadpanel}
  \end{subfigure}
  \hfill
  \begin{subfigure}[t]{0.32\textwidth}
    \centering
    \caption{Read isolation}
    \label{fig:load-buffering}
    \begin{threadpanel}
    \begin{threadtable}
      \threadhead
      \begin{threadcell}
        \begin{threadcode}
          a = x
          y = 1
          join T2
        \end{threadcode}
      \end{threadcell}
      &
      \begin{threadcell}
        \begin{threadcode}
          b = y
          x = 1
        \end{threadcode}
      \end{threadcell}
    \end{threadtable}
    \end{threadpanel}
  \end{subfigure}

  \medskip\smallskip



  \medskip
  {\footnotesize\itshape Isolated threads, communicating at their synchronisation}
  \par\smallskip
  
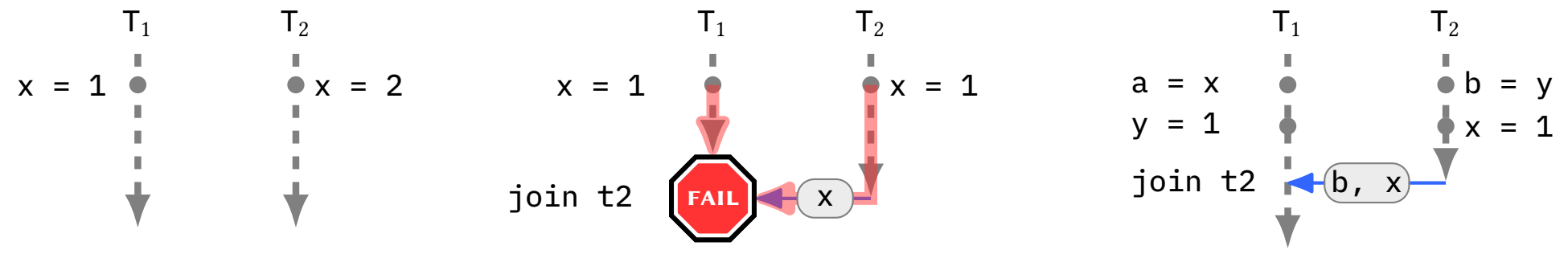
\begin{subfigure}[t]{0.32\textwidth}
    \SetLeftCodeShift{\CodeWidthOf{x = 1}}
    \SetRightCodeShift{3pt}
    \begin{tikzpicture}[common timeline styles, y=0.85cm]
      \TimelineColorSetup
      \newcommand{\onex}{0}
      \newcommand{\twox}{1.3}
      \node at (\onex,0.6) {T$_1$};
      \node at (\twox,0.6) {T$_2$};
      \DefineThreadEventMacros
      \useasboundingbox (-1.7,0.9) rectangle (2.1,-1.8);

      \ThreadEvent{t1}{-0}{t1x}{x = 1}
      \ThreadEvent{t2}{-0}{t2x}{x = 2}

      \draw[thread lane, ->] (\onex,0.3) -- (\onex,-1.4);
      \draw[thread lane, ->] (\twox,0.3) -- (\twox,-1.4);
    \end{tikzpicture}
  \end{subfigure}
  \hfill
  \begin{subfigure}[t]{0.32\textwidth}
    \SetLeftCodeShift{\CodeWidthOf{join T2}}
    \SetRightCodeShift{3pt}
    \begin{tikzpicture}[common timeline styles, y=0.85cm]
      \TimelineColorSetup
      \newcommand{\onex}{0}
      \newcommand{\twox}{1.3}
      \node at (\onex,0.6) {T$_1$};
      \node at (\twox,0.6) {T$_2$};
      \DefineThreadEventMacros
      \useasboundingbox (-1.7,0.9) rectangle (2.1,-1.8);

      \ThreadEvent{t1}{-0}{t1x}{x = 1}
      \ThreadSyncEventFrom{t1}{t1x}{-1.1}{t1join}{}
      \node[t1 code] at ($(t1join)+(-4mm,0)$) {join t2};
      \ConflictEvent{t1join}{t1conflict}

      \ThreadEvent{t2}{-0}{t2x}{x = 1}
      \ThreadSyncEvent{t2}{-1.1}{t2end}{}

      \draw[thread lane, ->] (\onex,0.3) -- (t1conflict);
      \draw[thread lane, ->] (\twox,0.3) -- (t2end);

      \draw[link oneway] (t2end) -- (t1conflict.east);

      \begin{scope}[transparency group, opacity=0.5]
        \draw[conflict] (t1x) -- (t1conflict.north);
        \draw[conflict] (t2x) -- (t2end) -- (t1conflict.east);
      \end{scope}
      \node[message] at ($(t2end)!0.4!(t1conflict.east)$) {x};

      \ConflictEvent{t1join}{t1conflicttop}
    \end{tikzpicture}
  \end{subfigure}
  \hfill
  \begin{subfigure}[t]{0.32\textwidth}
    \SetLeftCodeShift{\CodeWidthOf{join T2}}
    \SetRightCodeShift{3pt}
    \begin{tikzpicture}[common timeline styles, y=0.85cm]
      \TimelineColorSetup
      \newcommand{\onex}{0}
      \newcommand{\twox}{1.3}
      \node at (\onex,0.6) {T$_1$};
      \node at (\twox,0.6) {T$_2$};
      \DefineThreadEventMacros
      \useasboundingbox (-1.7,0.9) rectangle (2.1,-1.8);

      \ThreadEvent{t1}{-0}{t1a}{a = x}
      \ThreadEventFrom{t1}{t1a}{-0.4}{t1y}{y = 1}
      \ThreadSyncEventFrom{t1}{t1y}{-0.55}{t1join}{join t2}

      \ThreadEvent{t2}{-0}{t2b}{b = y}
      \ThreadEventFrom{t2}{t2b}{-0.4}{t2x}{x = 1}
      \ThreadSyncEvent{t2}{-0.95}{t2end}{}

      \draw[thread lane, ->] (\onex,0.3) -- (\onex,-1.6);
      \draw[thread lane, ->] (\twox,0.3) -- (t2end);

      \draw[link oneway] (t2end) -- (t1join);
      \node[message] at ($(t2end)!0.5!(t1join)$) {b, x};
    \end{tikzpicture}
  \end{subfigure}

  \setlength{\tabcolsep}{3.5pt}
  \footnotesize
  \setlength{\tabcolsep}{3pt}
  \newcolumntype{W}{>{\raggedright\arraybackslash}p{\dimexpr0.5\textwidth-15pt\relax}}
  \begin{tabular}{@{}l W @{\hspace{1em}} W}
    \toprule
     & \multicolumn{1}{c}{Shared Memory} & \multicolumn{1}{c}{\SyncasShare{}} \\
    \cmidrule(lr){2-2}\cmidrule(lr){3-3}
    \multirow{4}{*}{(\subref{fig:fork-no-join})}
      & \emph{Interpretation}: Both threads write \sm{shared} \c{x}, with different values. \sm{Nothing orders} them.
      & \emph{Interpretation}: Each thread writes its \sas{private} \c{x}. \sas{Nothing is communicated} between Threads~1 and 2. \\
    \addlinespace[0.45em]
      & \emph{Thread~1's Outcome}: \c{x} is 1 or 2, \sm{depending on} the schedule.
      & \emph{Thread~1's Outcome}: \c{x} is \sas{always} 1. \\
    \arrayrulecolor{black!10}\midrule\arrayrulecolor{black}
    \multirow{4}{*}{(\subref{fig:fork-join-write-write})}
      & \emph{Interpretation}: Both threads write \sm{shared} \c{x}, with the same value. \sm{Nothing orders} them.
      & \emph{Interpretation}: Each thread writes its \sas{private} \c{x}. The join \sas{communicates} Thread~2's update of \c{x} to Thread~1. \\
    \addlinespace[0.45em]
      & \emph{Thread~1's Outcome}: \c{x} is 1, but which write survives \sm{depends on} the schedule.
      & \emph{Thread~1's Outcome}: The incoming communication also updates \c{x}, so the join \sas{always} raises a conflict. \\
    \arrayrulecolor{black!10}\midrule\arrayrulecolor{black}
    \multirow{5}{*}{(\subref{fig:load-buffering})}
      & \emph{Interpretation}: Both \c{x} and \c{y} are \sm{shared}: one thread reads \c{x} and writes \c{y}, the other the reverse. \sm{Nothing orders} them.
      & \emph{Interpretation}: Each thread reads their \sas{private} \c{x} and \c{y}: one reads \c{x} and writes \c{y}, the other the reverse. The join \sas{communicates} Thread~2's update of \c{x} to Thread~1. \\
    \addlinespace[0.45em]
      & \emph{Thread~1's Outcome}: \c{a} and \c{b} each 0 or 1, \sm{depending on} the schedule and the compiler.
      & \emph{Thread~1's Outcome}: \c{a} and \c{b} are \sas{always} 0. \\
    \bottomrule
  \end{tabular}
  \vspace*{-1ex}
\caption{Fork-join programs, how each model reads them, and what Thread~1 is left with at the end of its own code.
  The two accounts differ in what the threads write to and in what relates them: \sm{shared} locations related by \sm{ordering}, or \sas{private} views related by \sas{communication}.
  The diagrams show what each program's synchronisation communicates; (a) has none, and the sign in (b) marks the thread that detects the conflict.}
  \Description[Three small fork-join programs used as examples.]{In (a) thread 1 writes one to x while thread 2 writes two to x, and the two never synchronise. In (b) thread 1 writes one to x and joins thread 2, while thread 2 writes one to x. In (c) thread 1 reads x into a, writes one to y, and joins thread 2, while thread 2 reads y into b and writes one to x. A table beneath gives, for each program, how each model reads it and what thread 1 is left with at the end of its own code. Under shared memory the threads write to shared locations and nothing orders them, so x is one or two depending on the schedule; x is one but which write survives depends on the schedule; and a and b are each zero or one depending on the schedule and the compiler. Under \SyncasShare{} each thread writes its own private copy: in the first nothing is communicated between the threads and x is always one; in the second the join communicates thread 2's update of x to thread 1 and there is always a conflict; in the third the threads write different variables, so the join always leaves a, b, and r2 all zero.}
  \label{fig:examples}
\end{figure}

We use the programs in \cref{fig:examples} to introduce the implications of isolation and explicit communication, and the programs in \cref{fig:lock-examples}  to present our two communication topologies.
We follow the standard form of ``litmus test'' from the memory model literature~\cite{AdveGharachorloo1996, Batty2011}.
Each program is expressed as two sequences of statements which may be interleaved, and possibly also reordered by a compiler under the assumption that each piece of mutable state is isolated to a single thread.
The variables \c{x}, \c{y}, \c{a} and \c{b} are global locations -- in a Java program, they might be fields of objects on the heap.

\subsubsection*{Figure 1(a): Threads are isolated in their writes}

Under \sm{Shared memory} this is a write--write race, but nothing marks it as such at run time.
Indeed, without a race detector, the program runs to completion with no sign that one thread's write was silently lost.
Under \sas{\SyncasShare{}} the programmer has provided no synchronisation.
As a result, neither thread's updates to its private view are ever communicated to the other.
Neither thread sees the other's change in memory.

\textbf{Insight:} Without the programmer providing synchronisation nothing a thread does is communicated, and what is never communicated cannot conflict.
This explicitly connects the programmer's interpretation of what data should be shared or isolated with their use of synchronisation.
However, what the runtime does at the end of a thread's lifetime matters, and we return to it in \cref{sec:syncasshare}.

\subsubsection*{Figure 1(b): Synchronisation is where updates are published and picked up}

Under \sm{Shared memory} the join is a wait in the execution, and says nothing about the data.
This write--write race is the textbook ``benign'' one: both threads write the same value, so nothing ever looks wrong.
However, one thread's write is lost with no  notification.
Under \sas{\SyncasShare{}} the join is a synchronisation event that causes an explicit communication of the two threads' updates to their private views.
Each thread compares \emph{what} the other has updated (not the values they wrote), so the two updates to \c{x} are detected as a conflict (despite both writing the same value).

\textbf{Insight:} A conflict can arise only at a synchronisation point, since that is where threads' updates are communicated to one another.
The thread executing the synchronisation operation is the one that detects the conflict, since threads receive updates only when they ask for them.
Whether two threads that write the same value should count as a conflict is our choice: we compare which locations each view updated, not the values written, so even this ``benign'' race is reported, while a value-sensitive check would accept it (\cf{} \cref{sec:what-is-a-conflict}).
However, neither choice affects correctly synchronised programs, which never bring two unordered writes together.

\subsubsection*{Figure 1(c): Threads are isolated in their reads}

Under \sm{Shared memory} this is a read--write race because the sources \c{x} and \c{y} are shared and updated.
Sequential consistency for global shared-memory admits three outcomes for \c{a} and \c{b}, and real compilers and hardware admit a fourth in which each read sees the write the other thread performs after it.
Under \sas{\SyncasShare{}} both reads happen before the synchronisation (join), so both read from their thread's private local view and get the original values for \c{x} and \c{y}.
The updates to \c{x} and \c{y} are communicated at the join, but as they do not conflict in \emph{what} was updated, there is no conflict and only one possible outcome.

\textbf{Insight:} That each thread's read returns the value in its own thread's view (\c{a=b=0}) follows from isolation alone, since a read cannot observe another thread's write until a synchronisation action communicates it.
However, having the join report no conflict in this case is a design choice.
We have chosen to track the locations of thread writes and not reads because an unreported read of this kind is harmless in a precise manner: it will always return a value permitted by JMM's happens-before consistency (\cf{} \cref{thm:hbc}).
Combined, we see that isolation removes the non-determinism of read--write races, but whether they should also be reported is a design choice.



\smallskip
\noindent
The above examples come from how thread isolation interacts with the structure of a fork/join program.
However, locks raise two different questions because they sit \emph{between} threads, rather than connecting them directly: \emph{what} is communicated and with \emph{whom}.
In \cref{fig:lock-examples}a we see that the only answer compatible with existing programs to \emph{what} to communicate is to communicate the thread's (full) view, \ie{} everything it has changed.
\Cref{fig:lock-examples}b shows that to \emph{whom} to communicate does not have a single answer, and we explore the choice of communication topologies in \cref{sec:syncasshare}.



\begin{figure}[t]
  \centering
  \captionsetup[subfigure]{position=top}
  {\footnotesize \textbf{Note:} at the start of all threads \c{x=m=0}}
  \par\smallskip
  \setlength{\threadpanelheight}{50pt}
  \begin{subfigure}[t]{0.40\textwidth}
    \centering
    \caption{Sharing scope}
    \label{fig:lock-scope}
    \begin{threadpanel}
    \begin{threadtable}
      \threadhead
      \begin{threadcell}
        \begin{threadcode}
          lock l
          assert(m == x)
          unlock l
        \end{threadcode}
      \end{threadcell}
      &
      \begin{threadcell}
        \begin{threadcode}
          x = 1
          lock l
          m = x
          unlock l
        \end{threadcode}
      \end{threadcell}
    \end{threadtable}
    \end{threadpanel}
  \end{subfigure}
  \hspace{2em}
  \begin{subfigure}[t]{0.40\textwidth}
    \centering
    \caption{Communication structure?}
    \label{fig:separate-lock-counter}
    \begin{threadpanel}
    \begin{threadtable}
      \threadhead
      \begin{threadcell}
        \begin{threadcode}
          lock l1
          m = m + 1
          unlock l1
        \end{threadcode}
      \end{threadcell}
      &
      \begin{threadcell}
        \begin{threadcode}
          lock l2
          m = m + 1
          unlock l2
        \end{threadcode}
      \end{threadcell}
    \end{threadtable}
    \end{threadpanel}
  \end{subfigure}

  \medskip\smallskip

  \medskip
  {\footnotesize\itshape Isolated threads, communicating at their synchronisation}
  \par\vskip-11pt
  \begin{subfigure}[t]{0.40\textwidth}
    \centering
    \SetLeftCodeShift{\CodeWidthOf{assert(m == x)}}
    \SetRightCodeShift{3pt}
    \begin{tikzpicture}[common timeline styles, y=0.8cm, baseline={(0,0.9)}]
      \TimelineColorSetup
      \newcommand{\onex}{0}
      \newcommand{\twox}{1.5}
      \node at (\onex,0.6) {T$_1$};
      \node at (\twox,0.6) {T$_2$};
      \DefineThreadEventMacros
      \useasboundingbox (-1.45,0.9) rectangle (2.95,-3.3);

      \ThreadEvent{t2}{-0.15}{t2x}{x = 1}
      \ThreadSyncEventFrom{t2}{t2x}{-0.4}{t2lock}{lock l}
      \ThreadEventFrom{t2}{t2lock}{-0.4}{t2m}{m = x}
      \ThreadSyncEventFrom{t2}{t2m}{-0.4}{t2unlock}{unlock l}

      \ThreadSyncEvent{t1}{-1.35}{t1lock}{lock l}
      \ThreadEventFrom{t1}{t1lock}{-0.4}{t1assert}{assert(m == x)}
      \ThreadSyncEventFrom{t1}{t1assert}{-0.4}{t1unlock}{unlock l}

      \draw[thread lane, ->] (\onex,0.3) -- (\onex,-2.5);
      \draw[thread lane, ->] (\twox,0.3) -- (\twox,-2.5);

      \draw[link oneway] (t2unlock) -- (t1lock);
      \node[message] at ($(t2unlock)!0.5!(t1lock)$) {x, m};
    \end{tikzpicture}
  \end{subfigure}
  \hspace{2em}
  \begin{subfigure}[t]{0.40\textwidth}
    \centering
    \SetLeftCodeShift{\CodeWidthOf{m = m + 1}}
    \SetRightCodeShift{3pt}
    \begin{tikzpicture}[common timeline styles, y=0.8cm, baseline={(0,0.9)}]
      \TimelineColorSetup
      \newcommand{\onex}{0}
      \newcommand{\twox}{1.8}
      \node at (\onex,0.6) {T$_1$};
      \node at (\twox,0.6) {T$_2$};
      \DefineThreadEventMacros
      \useasboundingbox (-1.3,0.9) rectangle (3.1,-2.0);

      \ThreadSyncEvent{t1}{-0.15}{t1lock}{lock l1}
      \ThreadEventFrom{t1}{t1lock}{-0.4}{t1inc}{m = m + 1}
      \ThreadSyncEventFrom{t1}{t1inc}{-0.4}{t1unlock}{unlock l1}

      \ThreadSyncEvent{t2}{-0.55}{t2lock}{lock l2}
      \ThreadEventFrom{t2}{t2lock}{-0.4}{t2inc}{m = m + 1}
      \ThreadSyncEventFrom{t2}{t2inc}{-0.4}{t2unlock}{unlock l2}

      \draw[thread lane, ->] (\onex,0.3) -- (\onex,-1.9);
      \draw[thread lane, ->] (\twox,0.3) -- (\twox,-1.9);

      \draw[open link] ($(t1lock)!0.3!(\twox,|-t1lock)$) -- (t1lock);
      \node[question] at ($(t1lock)!0.4!(\twox,|-t1lock)$) {};
      \draw[open link] ($(t2lock)!0.3!(\onex,|-t2lock)$) -- (t2lock);
      \node[question] at ($(t2lock)!0.4!(\onex,|-t2lock)$) {};

      \draw[open link] (t1unlock) -- ($(t1unlock)!0.3!(\twox,|-t1unlock)$);
      \node[question] at ($(t1unlock)!0.4!(\twox,|-t1unlock)$) {};
      \draw[open link] (t2unlock) -- ($(t2unlock)!0.3!(\onex,|-t2unlock)$);
      \node[question] at ($(t2unlock)!0.4!(\onex,|-t2unlock)$) {};

      \node[question] at ($(t1lock)!0.5!(t2unlock)$) {\Huge?};
    \end{tikzpicture}

    \smallskip
    {\footnotesize
      \begin{minipage}[c]{1.4em}
        \centering\color{communicationblue}\large\bfseries\sffamily ?
      \end{minipage}\hspace{0.35em}\begin{minipage}[c]{\dimexpr\linewidth-1.75em\relax}
        \raggedright\itshape Should lock actions communicate with every thread, or only with users of the same lock?
      \end{minipage}\par}
  \end{subfigure}

  \footnotesize
  \setlength{\tabcolsep}{3pt}
  \newcolumntype{V}{>{\raggedright\arraybackslash}p{\dimexpr0.5\textwidth-15pt\relax}}
  \begin{tabular}{@{}l V @{\hspace{1em}} V}
    \toprule
     & \multicolumn{1}{c}{Shared Memory} & \multicolumn{1}{c}{\SyncasShare{}} \\
    \cmidrule(lr){2-2}\cmidrule(lr){3-3}
    \multirow{4}{*}{(\subref{fig:lock-scope})}
      & \emph{Interpretation}: \c{x} and \c{m} are \sm{shared}. The lock \sm{orders} the accesses to \c{m}, but \c{x} is written before the lock is taken, so \sm{nothing orders} the accesses to \c{x}.
      & \emph{Interpretation}: Each thread has \sas{private} \c{x} and \c{m}. The lock \sas{communicates} Thread~2's whole view to Thread~1, or nothing at all. \\
    \addlinespace[0.45em]
      & \emph{Thread~1's Outcome}: The assertion can fail, \sm{depending on} which thread locks first.
      & \emph{Thread~1's Outcome}: The assertion \sas{always} holds. \\
    \arrayrulecolor{black!10}\midrule\arrayrulecolor{black}
    \multirow{5}{*}{(\subref{fig:separate-lock-counter})}
      & \emph{Interpretation}: \c{m} is \sm{shared}. The threads take different locks, so \sm{nothing orders} their updates.
      & \emph{Interpretation}: Each thread has a \sas{private} \c{m}. Each unlock \sas{communicates} to its own lock, and the threads hold no lock in common. \\
    \addlinespace[0.45em]
      & \emph{Thread~1's Outcome}: \c{m} is 1 or 2, \sm{depending on} the schedule.
      & \emph{Thread~1's Outcome}: \c{m} is 1 or 2, or the unlock reports a conflict, \sas{depending on} the schedule and \emph{how communication is structured} (\cref{sec:syncasshare}). \\
    \bottomrule
  \end{tabular}

  \caption{Lock programs, how each model reads them, and what Thread~1 is left with.
  Threads communicate with a lock, so they reach each other indirectly: in \subref{fig:lock-scope}
  they take the same lock and so communicate; in \subref{fig:separate-lock-counter} they take
  different locks, and in \cref{sec:syncasshare} we explore whether these threads still communicate.}
  \Description[Two small lock-based programs used as examples.]{In (a) thread 1 locks l, asserts that m equals x, and unlocks, while thread 2 writes one to x, locks l, assigns x to m, and unlocks. In (b) thread 1 locks l1, increments m, and unlocks l1, while thread 2 does the same with l2. A table beneath gives, for each program, how each model reads it and what thread 1 is left with. In (a) shared memory has x and m shared, and the lock orders the accesses to m but not those to x, which is written before the lock is taken, so the assertion can fail depending on which thread locks first; under \SyncasShare{} each thread has private copies and the lock communicates thread 2's whole view or nothing at all, so the assertion always holds. In (b) shared memory has m shared with the threads taking different locks, so nothing orders their updates and m is one or two depending on the schedule; under \SyncasShare{} each unlock communicates to its own lock and the threads hold no lock in common, so m is one or two depending on how communication is structured.}
  \label{fig:lock-examples}
\end{figure}

\subsubsection*{Figure 2(a): The synchronisation causes communication, not the ordering}

Under \sm{Shared memory} disciplined use of locks maintains isolation by controlling access ordering: if shared data is only touched under a lock, then only one thread works on it at a time.
Here that discipline is broken because Thread~1 accesses both \c{m} and \c{x} under the lock, while Thread~2 accesses \c{x} outside of the lock.
As a result, Thread~1 can lock first and read the initial \c{m} with either the initial \c{x} or Thread~2's new \c{x}.
Under \sas{\SyncasShare{}} the lock keeps the ordering role and gains another: it is when the thread's own updates are communicated outwards (a push) and others' updates inwards (a pull).
As a result, Thread~1 is guaranteed to get either both of Thread~2's updates (if Thread~2 locks first) or neither (if Thread~1 locks first).

\textbf{Insight:} An unlock communicates the thread's whole view, which includes writes made before the lock was taken.
This is the only choice consistent with today's semantics, as the JMM guarantees that Thread~1 sees \c{x=1} when Thread~2 locks first because the write to \c{x} happens before Thread~2's unlock.
This choice matters:
Choosing to have the unlock communicate only the writes made under the lock would give Thread~1 the new \c{m} and the old \c{x}, while shared memory permits the old \c{m} and the new \c{x}, when Thread~1 locks first.


\subsubsection*{Figure 2(b): Who communicates with whom?}
\begin{figure}[b]
  \begin{subfigure}[t]{0.51\textwidth}
    \SetLeftCodeShift{\CodeWidthOf{unlock(l1)}}
    \SetRightCodeShift{14pt}
    \begin{tikzpicture}[common timeline styles]
      \TimelineColorSetup

      \newcommand{\onex}{0}
      \newcommand{\gx}{1.3}
      \newcommand{\twox}{2.6}

      \node at (\onex,0.6) {T$_1$};
      \node at (\gx,0.6)   {g};
      \node at (\twox,0.6) {T$_2$};

      \DefineThreadEventMacros

      \ThreadSyncEvent{t1}{-0}{t1start}{<start>}
      \ThreadSyncEventFrom{t1}{t1start}{-0.8}{t1lock}{lock(l1)}
      \ThreadEventFrom{t1}{t1lock}{-0.4}{t1inc}{m = m + 1}
      \ThreadSyncEventFrom{t1}{t1inc}{-0.4}{t1unlock}{unlock(l1)}

      \ThreadSyncEvent{t2}{-0.1}{t2start}{<start>}
      \ThreadSyncEventFrom{t2}{t2start}{-1.1}{t2lock}{lock(l2)}
      \ThreadEventFrom{t2}{t2lock}{-0.6}{t2inc}{m = m + 1}
      \ThreadSyncEventFrom{t2}{t2inc}{-1.0}{t2unlock}{unlock(l2)}
      \ConflictEvent{t2unlock}{t2conflict}

      \draw[thread lane, ->] (t1start) -- (\onex,-4.2);
      \draw[shared lane, ->] (\gx,0.3)   -- (\gx,-4.2);
      \draw[thread lane, ->] (t2start) -- (t2conflict);

      \draw[link oneway] (\gx,|-t1start) -- node[message] {m} (t1start);
      \draw[link oneway] (\gx,|-t1lock) -- (t1lock);
      \draw[link oneway] (\gx,|-t1unlock) to[looseness=.5, out=140, in=30] (t1unlock);
      \draw[link oneway] (t1unlock) to[looseness=.5, out=320, in=210]
        coordinate[pos=0.5] (pushmsg) (\gx,|-t1unlock);

      \draw[link oneway] (\gx,|-t2start) -- node[message] {m} (t2start);
      \draw[link oneway] (\gx,|-t2lock) -- (t2lock);
      \draw[link oneway] (\gx,|-t2unlock) to[looseness=.5, out=40, in=150]
        coordinate[pos=0.45] (pullmsg) (t2conflict.west);

      \begin{scope}[transparency group, opacity=0.5]
        \draw[conflict] (t1inc) --  (t1unlock) to[bend right, looseness=.5, out=320, in=210] (\gx,|-t1unlock) -- (\gx, |-t2unlock) to[looseness=.5, out=40, in=150] (t2conflict.west);
        \draw[conflict] (t2inc) -- (t2conflict);
      \end{scope}
      \node[msg] at (pushmsg) {m};
      \node[msg] at (pullmsg) {m};
    \end{tikzpicture}
    \caption{A conflicting execution.}
    \label{fig:linear-counter-conflict}
  \end{subfigure}
  \hfill
  \begin{subfigure}[t]{0.47\textwidth}
    \SetLeftCodeShift{\CodeWidthOf{unlock(l1)}}
    \begin{tikzpicture}[common timeline styles]
      \TimelineColorSetup

      \newcommand{\onex}{0}
      \newcommand{\gx}{1.3}
      \newcommand{\twox}{2.6}

      \node at (\onex,0.6) {T$_1$};
      \node at (\gx,0.6)   {g};
      \node at (\twox,0.6) {T$_2$};

      \DefineThreadEventMacros

      \ThreadSyncEvent{t1}{-0}{t1start}{<start>}
      \ThreadSyncEventFrom{t1}{t1start}{-0.8}{t1lock}{lock(l1)}
      \ThreadEventFrom{t1}{t1lock}{-0.4}{t1inc}{m = m + 1}
      \ThreadSyncEventFrom{t1}{t1inc}{-0.4}{t1unlock}{unlock(l1)}

      \ThreadSyncEvent{t2}{-0.1}{t2start}{<start>}
      \ThreadSyncEventFrom{t2}{t2start}{-1.8}{t2lock}{lock(l2)}
      \ThreadEventFrom{t2}{t2lock}{-0.4}{t2inc}{m = m + 1}
      \ThreadSyncEventFrom{t2}{t2inc}{-0.4}{t2unlock}{unlock(l2)}
      \ThreadSyncEventFrom{t2}{t2unlock}{-0.8}{t2end}{<end>}

      \draw[thread lane, ->] (t1start) -- (\onex,-4.2);
      \draw[thread lane, ->] (t2start) -- (t2end);

      \draw[shared lane] (\gx,0.3) -- ($(\gx,|-t1unlock)+(0,4pt)$);
      \draw[shared lane] ($(\gx,|-t1unlock)-(0,4pt)$) -- ($(\gx,|-t2unlock)+(0,4pt)$);
      \draw[shared lane, ->] ($(\gx,|-t2unlock)-(0,4pt)$) -- (\gx,-4.2);

      \draw[link oneway] (\gx,|-t1start) -- node[message] {m} (t1start);
      \draw[link oneway] (\gx,|-t1lock) -- (t1lock);
      \PullPushMsg{(t1unlock)}{(\gx,|-t1unlock)}{m}

      \draw[link oneway] (\gx,|-t2start) -- node[message] {m} (t2start);
      \draw[link oneway] (\gx,|-t2lock) -- node[message] {m} (t2lock);
      \PullPushMsg{(t2unlock)}{(\gx,|-t2unlock)}{m}
      \PullPush{(t2end)}{(\gx,|-t2end)}
    \end{tikzpicture}
    \caption{A non-conflicting execution.}
    \label{fig:linear-counter-no-conflict}
  \end{subfigure}
  \caption{Executions of the separate-lock counter under \LinearW{} \SyncasShare{} communication. The central lane $g$ is the shared synchronisation object; $m_1$ and $m_2$ are the private views of the counter in T$_1$ and T$_2$.}
  \Description[Linear Sync-as-Share executions of the separate-lock counter.]{The first timeline shows two threads incrementing private views and a later conflict when the second update meets the first through the shared synchronisation object. The second timeline shows the first update being communicated through the shared synchronisation object to the second thread before its increment.}
  \label{fig:linear-counter-executions}
\end{figure}
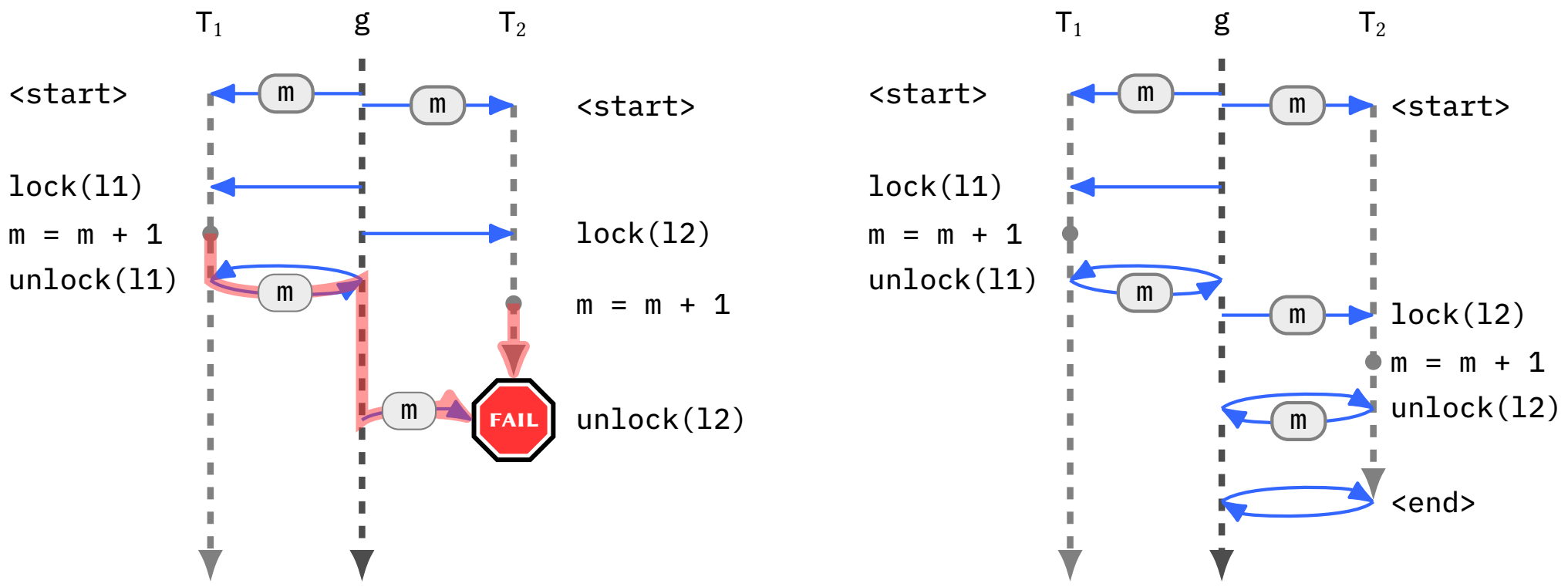

Under \sm{Shared memory} every access to \c{m} is under a lock, yet it is still a write--write race as the programmer failed to consistently use the same lock for the same data.
Under \sas{\SyncasShare{}} the critical sections are also not ordered.
However, whether the two threads' updates ever meet depends on \emph{the communication topology}.
If synchronisation actions communicate through a global view, the choice of lock is irrelevant and the updates meet.
When the critical sections do not overlap, the second thread picks up the first thread's increment when its lock communicates their views and \c{m} is correctly incremented.
When they do overlap, the second thread to unlock sees the first thread's update pushed to the global view and reports a conflict.
However, if updates are only communicated between threads that operate on the same lock, the updates will not meet, and the conflict will only be detected later if something else, such as a join, brings the views together.
We explore these design choices in \cref{sec:syncasshare}, and return to this program in \cref{fig:linear-counter-executions,fig:branching-counter-execution}.
\TightenPar{1}

\textbf{Insight:} The topology choice does not matter for correctly synchronised programs, which will behave exactly as under the JMM (\cf{} \cref{thm:equiv}).
However, it does matter for programs like this one, where locking is not used correctly, and decides whether and where that incorrectness surfaces.
A single global view ``leaks'' updates to threads that synchronise through different locks, so the schedule decides whether two of them conflict, much as it decides whether a shared-memory program exhibits a race.
A view per lock follows the synchronisation the program performs, so only threads that synchronise with one another can conflict, whatever the schedule.

\medskip

\SyncasShare{} removes the data races from every one of these programs, by starting from per-thread isolation and extending existing ordering synchronisation to communicate explicitly.
This suits managed languages such as Java, which are expected to provide the same behaviour on a range of hardware memory models, since isolation no longer depends on how any one of them orders accesses.
In contrast, a low-level language deliberately exposes that ordering to the programmer.
Portability is therefore an additional reason to instantiate \SyncasShare{} for Java.

\section{Communication Topologies}
\label{sec:syncasshare}

%
In this section we present and motivate the two communication topologies we have identified for \SyncasShare{}.
We start by recapitulating the terminology used in this paper.
A \emph{private view} is a view of memory belonging to a single thread; ordinary reads and writes operate only on that view.
A \emph{synchronisation object} is what two threads communicate through. In Java these are monitors, volatile fields, and threads.
A \emph{synchronisation action} is an operation that both orders thread execution and communicates updates between a thread's private view and the synchronisation object it names.
A \emph{conflict} arises when a synchronisation action brings together two views that have both updated the same location, regardless of the values written.
A conflict is reported as an exception thrown in the thread that performs the synchronisation action.

The two locks in \cref{fig:separate-lock-counter} gave rise to the question of communication topology: to whom does a thread push out its updates, and who can pull them in?
Java's memory model orders synchronisation actions in two ways~\cite[\S17.4.4]{JMM}, and each suggests a topology.
\emph{Synchronisation order} is a total order over all the synchronisation actions of an execution.
\emph{Synchronizes-with} is partial and relates an action only to actions on the same synchronisation object.
For example, an unlock of a monitor is related to every subsequent lock of that monitor.
Under \SyncasShare{} those actions are where threads communicate, so either order can be taken as saying which threads communicate with which.
\TightenPar{1}

The \emph{\LinearW{}} topology (\cf{} \cref{sec:linear}) comes from synchronisation order.
Every action is ordered against every other, so every thread's updates reach every other thread, accumulating in a single global view.
The history is a straight line.
The \emph{\BranchingW{}} topology (\cf{} \cref{sec:branching}) comes from synchronizes-with.
An action is related only to actions on the same object, so communication follows the objects the program names and each object carries a view of its own.
The history branches and rejoins instead of running in a straight line.

Java builds happens-before from synchronizes-with, so communication in \BranchingW{} happens exactly where Java requires it.
Communication in \LinearW{} also happens when threads synchronise, but reaches wider than just the involved threads.
Java permits this wider communication.
In shared memory, communication already happens at every ordinary read and write, with no synchronisation at all.
In \SyncasShare{} it happens only at synchronisation actions.

There is a trade-off in these models.
\BranchingW{} better represents the synchronizes-with order, and therefore the happens-before order, but its branches must be reconciled wherever they join, and the branching history makes that check more complicated.
\LinearW{} is a coarser over-approximation of synchronizes-with, and the straight-line history makes those reconciliation points simpler.

\Cref{tab:syncasshare-java-intuition} shows the meaning of Java's synchronisation actions in the two topologies.
Each of those actions either releases or acquires~\cite[\S17.4.4]{JMM}.
Under \SyncasShare{} a release pushes and an acquire pulls.
\Push{t}{x} pushes thread $t$'s updates to the synchronisation object $x$.
In \LinearW{}, $x$ is always the global view $g$.
In \BranchingW{}, $x$ can be another thread, a lock, or a volatile field.
\Pull{t}{x} pulls all outstanding updates from $x$, \ie{} those pushed to $x$ since $t$'s last pull from $x$.
A pull fails if it detects a conflict.
\Pullpush{t}{x} atomically combines the two, and aborts before the push if the pull fails, so no conflicting view is ever pushed out.
In \LinearW{} a release pulls first as well, since its push has to be checked against the global view.
\TightenPar{1}

\begin{table}[hbt]
  \caption{Java illustration of \LinearW{} and \BranchingW{} \SyncasShare{} communication.}
  \Description[Table illustrating Linear and Branching Sync-as-Share communication with Java actions.]{Thread start, thread join, monitor lock, monitor unlock, volatile reads, and volatile writes are mapped to communication through either a single shared synchronisation object or the synchronisation object named by the program.}
  \label{tab:syncasshare-java-intuition}
  \begingroup
  \centering
  \begin{minipage}[t]{0.49\columnwidth}
    \centering
    \begin{tabular}[t]{ccll}
      \toprule
      \multicolumn{2}{c}{Sync. Action} & \multicolumn{2}{c}{Reinterpretation}                  \\
      \cmidrule(lr){1-2}\cmidrule(lr){3-4}
      Object                    & Action           & \LinearW{}                     & \BranchingW{}   \\
      \cmidrule(lr){1-1}\cmidrule(lr){2-2}\cmidrule(lr){3-3}\cmidrule(lr){4-4}
      \multirow{2}{*}{thread}   & spawn $t'$   & \Pullpush{t}{g}   & \Push{t}{t'} \\
                                    & join $t'$    & \Pull{t}{g}        & \Pull{t}{t'} \\
      \cmidrule(lr){1-4}
      \multirow{2}{*}{lock}         & lock $l$     & \Pull{t}{g}        & \Pull{t}{l}  \\
                                    & unlock $l$   & \Pullpush{t}{g}   & \Push{t}{l}  \\
      \cmidrule(lr){1-4}
      \multirow{2}{*}{volatile}     & read $v$     & \Pull{t}{g}        & \Pull{t}{v}  \\
                                    & write $v$    & \Pullpush{t}{g}   & \Pullpush{t}{v}  \\
      \bottomrule
    \end{tabular}
  \end{minipage}\hfill
  \begin{minipage}[t]{0.49\columnwidth}
    \centering
    \begin{tabular}[t]{clc}
      \toprule
      Sync. Action & \multicolumn{2}{c}{Reinterpretation}   \\
      \cmidrule(lr){1-2}\cmidrule(lr){2-3}
                   & \LinearW{}      & \BranchingW{}   \\
      \cmidrule(lr){1-1}\cmidrule(lr){2-2}\cmidrule(lr){3-3}
      start        & \Pull{t}{g}     & --- \\
      end          & \Pullpush{t}{g} & --- \\
      \bottomrule
    \end{tabular}

    \medskip\medskip

    \begin{tabular}[t]{ll}
      \Pull{t}{x} & \parbox[t]{0.59\columnwidth}{\raggedright$t$ pulls updates from $x$, checking~for conflicts.} \\[1.50em]
      \Push{t}{x} & \parbox[t]{0.59\columnwidth}{\raggedright$t$ pushes updates to $x$.} \\[0.5em]
      \Pullpush{t}{x} & \parbox[t]{0.59\columnwidth}{\raggedright$t$ atomically pulls from and, if~conflict-free, pushes to $x$.} \\
    \end{tabular}
  \end{minipage}
  \endgroup
  \smallskip
\end{table}

The right table shows the meaning of the synthetic synchronisation actions \emph{start} and \emph{end} which constitute the first and last action of a thread, respectively.
In the \LinearW{} model, a thread fetches all updates from the global view $g$ at its start, and, at its end, pulls in all outstanding updates, reconciles them and then pushes back a merged view of its state to $g$ (or aborts if the pull detected a conflict).
In \BranchingW{}, both synthetic actions are no-ops.
This is because the spawn action pushes the parent thread's view to the child when the child is spawned, and because a join in \BranchingW{} pulls directly from the thread instead of from $g$.
A thread that is never joined therefore communicates nothing at its end, which is what synchronizes-with gives, since no join means its final action is related to nothing.
The volatile write is the one place \BranchingW{} departs from synchronizes-with, since it pulls as well as pushes where the relation gives a write no incoming edge.
We discuss this part of the design space in \cref{sec:volatile}.

\subsection{\LinearW{} communication}
\label{sec:linear}

To illustrate the effect of the \LinearW{} topology, we examine the separate-lock counter example from \cref{fig:separate-lock-counter} and present two executions in \cref{fig:linear-counter-executions}.
Whether a conflict is detected depends on the ordering.
\textbf{In the conflicting execution (left):} Thread~1 starts and locks $l_1$, pulling from $g$, sees the initial value of the counter, and increments its private view of the counter.
Thread~2 starts and locks $l_2$, pulling from $g$ and also sees the initial value of the counter, before incrementing its private view.
Thread~1 then unlocks $l_1$, pulling from $g$ (finding no updates) and pushing its updated value of the counter ($m$) to $g$.
When Thread~2 unlocks $l_2$, it pulls from $g$ and discovers the update from Thread~1, but as it has already updated its own private view of $m$ in the meantime, this constitutes a conflict and an exception is raised.
\textbf{In the non-conflicting execution (right):}
both threads start as before, each pulling the initial value of the counter from $g$ into their private views.
Thread~1 locks $l_1$, increments its private view, and unlocks $l_1$, pushing its update to $g$.
Thread~2 then locks $l_2$, pulling from $g$, which now incorporates Thread~1's update.
Thread~2 then increments its private view and unlocks $l_2$, pushing its updates to $g$.
The execution is non-conflicting as Thread~2's final push does not see a conflict because it had earlier pulled Thread~1's updates from $g$.

The non-conflicting nature of the second execution is the result of \emph{information leakage}: the synchronisation actions on $l_1$ and $l_2$ are unrelated, yet the updates to $m$ are communicated between the threads through the global $g$ that \LinearW{} uses.
The issue of leakage further motivates the more directed communication topology of the \BranchingW{} topology.

\begin{figure}[t]
  \begin{subfigure}[t]{0.48\textwidth}
        \centering
    \SetLeftCodeShift{\CodeWidthOf{unlock(l1)}}
    \begin{tikzpicture}[common timeline styles]
      \TimelineColorSetup

      \newcommand{\onex}{0}
      \newcommand{\lonex}{1.1}
      \newcommand{\ltwoex}{2.2}
      \newcommand{\twox}{3.3}

      \node at (\onex,0.6) {T$_1$};
      \node at (\lonex,0.6) {L$_1$};
      \node at (\ltwoex,0.6) {L$_2$};
      \node at (\twox,0.6) {T$_2$};

      \DefineThreadEventMacros

      \ThreadSyncEvent{t1}{-0.2}{t1lock}{lock(l1)}
      \ThreadEventFrom{t1}{t1lock}{-0.4}{t1inc}{m = m + 1}
      \ThreadSyncEventFrom{t1}{t1inc}{-0.4}{t1unlock}{unlock(l1)}

      \ThreadSyncEvent{t2}{-0.6}{t2lock}{lock(l2)}
      \ThreadEventFrom{t2}{t2lock}{-0.6}{t2inc}{m = m + 1}
      \ThreadSyncEventFrom{t2}{t2inc}{-0.6}{t2unlock}{unlock(l2)}

      \draw[thread lane, ->] (\onex,0.3) -- (\onex,-3.5);
      \draw[shared lane, ->] (\lonex,0.3)   -- (\lonex,-3.5);
      \draw[thread lane, ->] (\twox,0.3) -- (\twox,-3.5);
      \draw[shared lane, ->] (\ltwoex,0.3) -- (\ltwoex,-3.5);

      \draw[link oneway] (\lonex,|-t1lock) -- (t1lock);
      \draw[link oneway] (t1unlock) -- node[message] {m} (\lonex,|-t1unlock);

      \draw[link oneway] (\ltwoex,|-t2lock) -- (t2lock);
      \draw[link oneway] (t2unlock) -- node[message] {m} (\ltwoex,|-t2unlock);

    \end{tikzpicture}
    \caption{The separate-lock counter under \BranchingW{} \SyncasShare{} communication. The lock actions communicate through separate synchronisation objects and so do not conflict.}
    \Description[Branching Sync-as-Share execution of the separate-lock counter.]{Thread one publishes its counter update through the first monitor and thread two publishes its update through the second monitor. A later join attempts to combine the two updates and reaches a conflict.}
    \label{fig:branching-counter-execution-nonconflict}
  \end{subfigure}
  \hfill
  \begin{subfigure}[t]{0.48\textwidth}
    \centering
    \SetLeftCodeShift{\CodeWidthOf{unlock(l1)}}
    \begin{tikzpicture}[common timeline styles]
      \TimelineColorSetup

      \newcommand{\onex}{0}
      \newcommand{\lonex}{1.1}
      \newcommand{\ltwoex}{2.2}
      \newcommand{\twox}{3.3}

      \node at (\onex,0.6) {T$_1$};
      \node at (\lonex,0.6) {L$_1$};
      \node at (\ltwoex,0.6) {L$_2$};
      \node at (\twox,0.6) {T$_2$};

      \DefineThreadEventMacros

      \ThreadSyncEvent{t1}{-0.2}{t1lock}{lock(l1)}
      \ThreadEventFrom{t1}{t1lock}{-0.4}{t1inc}{m = m + 1}
      \ThreadSyncEventFrom{t1}{t1inc}{-0.4}{t1unlock}{unlock(l1)}
      \ThreadSyncEventFrom{t1}{t1unlock}{-1.3}{t1join}{join(t2)}
      \ConflictEvent{t1join}{t1conflict}

      \ThreadSyncEvent{t2}{-0.6}{t2lock}{lock(l2)}
      \ThreadEventFrom{t2}{t2lock}{-0.6}{t2inc}{m = m + 1}
      \ThreadSyncEventFrom{t2}{t2inc}{-0.6}{t2unlock}{unlock(l2)}

      \draw[thread lane, ->] (\onex,0.3) -- (t1conflict);
      \draw[shared lane, ->] (\lonex,0.3)   -- (\lonex,-3.5);
      \draw[thread lane, ->] (\twox,0.3) -- (\twox,|-t1join);
      \draw[shared lane, ->] (\ltwoex,0.3) -- (\ltwoex,-3.5);

      \draw[link oneway] (\lonex,|-t1lock) -- (t1lock);
      \draw[link oneway] (t1unlock) -- node[message] {m} (\lonex,|-t1unlock);
      \draw[link oneway] (\twox,|-t1join) -- coordinate (joinmsg) (t1conflict);

      \draw[link oneway] (\ltwoex,|-t2lock) -- (t2lock);
      \draw[link oneway] (t2unlock) -- node[message] {m} (\ltwoex,|-t2unlock);

      \begin{scope}[transparency group, opacity=0.5]
        \draw[conflict] (t1inc) -- (t1conflict);
        \draw[conflict] (t2inc) -- (\twox,|-t1join) -- (t1conflict);
      \end{scope}
      \node[msg] at (joinmsg) {m};
    \end{tikzpicture}
    \caption{Adding a join to the previous execution causes a conflict. The join attempts to combine the two updates, which were communicated through separate synchronisation objects and so cannot be ordered.}
    \Description[Branching Sync-as-Share execution of the separate-lock counter.]{Thread one publishes its counter update through the first monitor and thread two publishes its update through the second monitor. A later join attempts to combine the two updates and reaches a conflict.}
    \label{fig:branching-counter-execution-conflict}
  \end{subfigure}
  \caption{Executions of the separate-lock counter under \BranchingW{} \SyncasShare{} communication.}
  \label{fig:branching-counter-execution}
\end{figure}
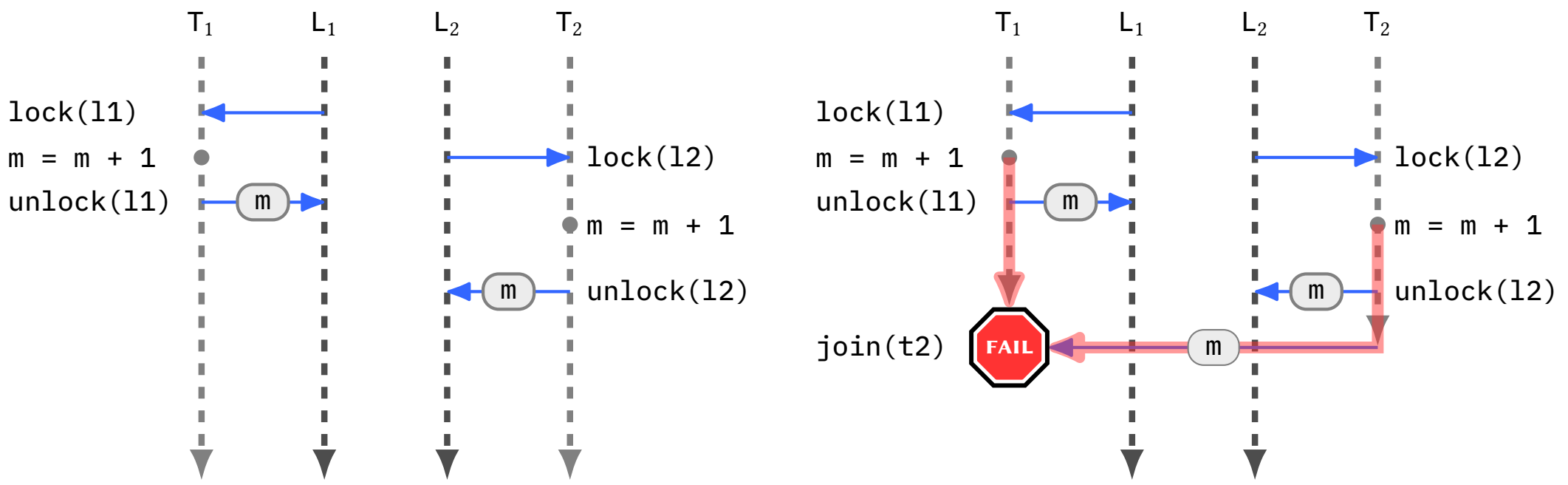

\subsection{\BranchingW{} communication}
\label{sec:branching}

In \BranchingW{}, the synchronisation actions on $l_1$ and $l_2$ communicate through separate synchronisation objects, so they do not leak information about updates to $m$ between the threads and no conflict can be observed.
This can be seen in \cref{fig:branching-counter-execution} left, which shows the same interleaving that caused a conflict under \LinearW{}, but here the pushes are to two separate synchronisation objects ($l_1$ and $l_2$) instead of a shared global one ($g$).
The JMM requires only this much communication for this example, but it permits more, since a read is allowed to return the value of an unordered write from another thread.
In \BranchingW{} the read can only return the value in the reading thread's own view, which is one of the options the JMM allows (\cref{thm:hbc}).

If Thread~1 joins Thread~2, Thread~1 will pull in updates from Thread~2.
This means the conflict will necessarily surface since the two threads' views each contain separate updates to $m$.
\Cref{fig:branching-counter-execution} right shows how a join in Thread~1 forces a synchronisation between two threads' views, leading to a conflict detection.
Indeed, no schedule can establish a coherent order for them before the join detects a conflict, which is the benefit of the reduced information leakage of \BranchingW{} over \LinearW{}.

\TightenPar{1}


\paragraph{Information transfer through synchronisation.}
\BranchingW{} prevents updates from reaching a thread through synchronisation objects it never explicitly names, but the communication itself is un-scoped: a synchronisation action carries all updates a thread has accumulated.
The consequence of this is visible in nested locking, as the inner locks will communicate all updates from the thread's private view, including those done under the outer lock.
For example, if a thread locks $l_1$, writes $x$, locks $l_2$, writes $y$, and unlocks $l_2$, then $l_2$ receives a view containing both $x$ and $y$, not just $y$.
A later thread acquiring $l_2$ therefore sees not just $y$ but also $x$, which was written before $l_2$ was even locked.
This behavior is important to satisfy the release-acquire guarantee that correctly synchronised Java programs already rely on.
\SyncasShare{} produces this naturally because synchronisation communicates a view, not a selection of writes\footnote{Providing a ``hierarchical'' scoping, in which only writes under the current synchronisation object are communicated, would limit the state transfer further, but be incompatible with what the JMM requires today.}.

\section{Realizing \SyncasShare{} via Version-Controlled memory}
\label{sec:vcm}
\label{sec:formal-semantics}

We make the two \SyncasShare{} communication topologies precise by formalising their semantics in what we call \emph{version-controlled memory}.
In version-controlled memory, each thread maintains a diff for uncommitted writes and a commit identifier recording its position in the shared history; synchronisation actions advance or extend that position.
The version information is what allows the model to detect conflicts: two views that have made incompatible independent updates cannot be coherently combined.

In \LinearW{}, all synchronisation actions share a single global version history $g$; in \BranchingW{}, threads and synchronisation objects each track a commit in a shared commit graph.
The \LinearW{} history resembles the linear revision history of Subversion or multi-version concurrency control (MVCC), while the \BranchingW{} commit graph resembles the commit graph of a distributed version control system such as Git.
\cref{tab:vcm-operation-rules} defines pull, push, and pull-push for both topologies.

\begin{table}[t]
  \centering
  \small
  \begingroup
  \settowidth{\syncarrowwidth}{$\longleftarrow$}
  \renewcommand{\pull}[2]{\ensuremath{#1 \mathrel{\makebox[\syncarrowwidth][l]{$\longleftarrow$}} #2}}
  \renewcommand{\push}[2]{\ensuremath{#1 \mathrel{\makebox[\syncarrowwidth][l]{$\longrightarrow$}} #2}}
  \newcommand{\pullpush}[2]{
    \ensuremath{#1%
      \mathrel{%
        \makebox[\syncarrowwidth][l]{%
          \tikz[baseline=-0.5ex, x=\syncarrowwidth]{
            \draw[{Classical TikZ Rightarrow}-,bend left=40, line width=0.7pt] (0,0) to (1,0.02);
            \draw[{Classical TikZ Rightarrow}-,bend left=40, line width=0.7pt] (1,0) to (0,-0.02);
          }
        }
      }%
    #2}
  }
  \caption{Version-controlled realisation of pull, push, and pull-push. $x$'s commit denotes the commit $x$ currently points to; B-Pull and B-Push each cover a family of rules (see \cref{sec:branchingsem}).}
  \label{tab:vcm-operation-rules}
  \begin{tabular}{lp{0.36\columnwidth}p{0.42\columnwidth}}
    \toprule
    Operation & \LinearW{} ($x = g$, version history) & \BranchingW{} (commit graph) \\
    \midrule
    \pull{t}{x} &
      \textbf{L-Pull}: if $t$'s staged writes do not conflict with updates to $g$ since $t$'s last pull, advance $t$'s commit to $g$'s latest. &
      \textbf{B-Pull}: if merging $t$'s commit with $x$'s commit is conflict-free, extend the graph with a merge commit as a child of both; $t$ advances to it. \\[0.5em]
    \push{t}{x} &
      \textit{(not applicable in \LinearW{})} &
      \textbf{B-Push}: no conflict check; create a new commit from $t$'s staged writes as a child of $t$'s current commit; advance $x$'s commit to it. \\[0.5em]
    \pullpush{t}{x} &
      \textbf{L-PullPush}: atomically, as L-Pull then append $t$'s staged writes to $g$ as a new commit; $t$ and $g$ both advance to it. &
      \textbf{B-PullPush}: atomically, as B-Pull from $x$'s commit then B-Push to $x$'s commit \textit{(not in the paper)}. \\
    \bottomrule
  \end{tabular}
  \endgroup
\end{table}

\begin{figure}
\begin{subfigure}{\textwidth}
  \begin{mathpar}
    v \in \Val \and \mathit{exc} \in \Exc \and r \in \Val \cup \Exc \and e \in \Exp \supseteq \left(\Val \cup \Exc\right)\\
    t \in \TID \and x \in \Var \and \ell \in \Lock \and \Versioned{c \in \CommitID}\\
    \alpha \in \{\kw{read}\ x,\ \kw{write}\ x\ v,\ \kw{spawn}\ e,\ \kw{join}\ t,\ \kw{lock}\ \ell,\ \kw{unlock}\ \ell\} \and K \in \left(\Val \cup \{\unit\} \cup \Exc \cup \TID\right) \rightarrow \Exp
  \end{mathpar}

  We assume a relation $e \emit{\alpha} K$ that steps $e$ to $K$, emitting effect $\alpha$.

  \[
    \begin{array}{cc}
    \begin{array}{rcll}
      \mathit{ts} & ::= & \new ~|~ \running ~|~ \terminated & \textit{(thread state)}\\
      \Versioned{\delta} & \in & \Versioned{\Var \rightharpoonup Val} & \textit{(diff)}\\
      \mathit{thread} & ::= & (\mathit{ts}, e, \Versioned{\delta, c})\\
      T & \in & t \rightharpoonup \mathit{thread}\\
      \Heap & ::= & \textit{(see \cref{fig:heap-semantics})} \\
    \end{array}&
    \begin{array}{rcll}
      \mathit{ls} & ::= & \kw{free} ~|~ \kw{held}(t) & \textit{(lock state)}\\
      \mathit{lock} & ::= & (\mathit{ls}, \Versioned{c})\\
      L & \in & \ell \rightarrow \mathit{lock} \\
      \mathit{volatile} & ::= & (\mathit{val}, \Versioned{c})\\
      V & \in & \Var \rightarrow \mathit{volatile} \\
    \end{array}
    \end{array}
  \]

  \caption{Syntax and sequential semantics.}
  \Description[Syntax and sequential semantics.]{The figure gives the domains of values, exceptions, results, expressions, thread identifiers, variables, locks, effects, continuations, thread states, threads, lock states, locks, and heaps.}
  \label{fig:sequential}
\end{subfigure}

\begin{subfigure}{\textwidth}
{ \small
\begin{mathpar}
  \inferrule*[right=Start]
  {T(t) = (\new, e, \Versioned{\delta, c}) \\\\ \Versioned{\Init(H,c) = c'}}
  {\Conf{T}{\Heap}{\Locks} \step \Conf{T\upd{t}{(\running, e, \Versioned{\delta, c'})}}{\Heap}{\Locks}}

  \inferrule*[right=End]
  {T(t) = (\running, r, \Versioned{\delta, c}) \\\\ \Versioned{\Conf{T}{\Heap}{\Locks} \commit{t} \Conf{T'}{\Heap'}{\Locks}}}
  {\Conf{T}{\Heap}{\Locks} \step \Conf{T'\upd[ts]{t}{\terminated}}{\Heap'}{\Locks}}

  \inferrule*[right=Read]
  {T(t) = (\running, e, \Versioned{\delta, c}) \and e \emit{\kw{read}\ x} K\\\\
    v = \Read(\Heap,\Versioned{\delta, c}, x)}
  {\Conf{T}{\Heap}{\Locks} \step \Conf{T\upd[e]{t}{\apply{K}{v}}}{\Heap}{\Locks}}

  \inferrule*[right=Write]
  {T(t) = (\running, e, \Versioned{\delta, c}) \and e \emit{\kw{write}\ x\ v} K \\\\
    (\Heap', \Versioned{\delta'}) = \Write(\Heap, \Versioned{\delta}, x, v) \\\\
    T' = T\upd{t}{(\running, \apply{K}{\unit}, \Versioned{\delta', c})} }
  {\Conf{T}{\Heap}{\Locks} \step \Conf{T'}{\Heap'}{\Locks}}

  \inferrule*[right=Spawn]
  {T(t) = (\running, e, \Versioned{\delta, c}) \and e \emit{\kw{spawn}\ e_s} K \and
    T' = T\upd[e]{t}{\apply{K}{t'}}\upd{t'}{(\new, e_s, \Versioned{\emptyset, 0})} \\
    \Versioned{\Conf{T'}{\Heap}{\Locks} \push{t, t'} \Conf{T''}{\Heap'}{\Locks}} \and t' \notin \dom(T)}
  {\Conf{T}{\Heap}{\Locks} \step \Conf{T''}{\Heap'}{\Locks}}

  \inferrule*[right=Join]
  {T(t) = (\running, e, \Versioned{\delta, c}) \and e \emit{\kw{join}\ t'} K \and
   T(t') = (\terminated, r, \Versioned{\_, \_}) \and
   \Versioned{\Conf{T}{\Heap}{\Locks} \pull{t, t'} \Conf{T'}{\Heap'}{\Locks}}}
  {\Conf{T}{\Heap}{\Locks} \step \Conf{T'\upd[e]{t}{\apply{K}{r}}}{\Heap'}{\Locks}}

  \inferrule*[right=Lock]
  {T(t) = (\running, e, \Versioned{\delta, c}) \and e \emit{\kw{lock}\ \ell} K \and \Locks(\ell).ls = \kw{free} \and \Versioned{\Conf{T}{\Heap}{\Locks} \pull{t, \ell} \Conf{T'}{\Heap'}{\Locks}}}
  {\Conf{T}{\Heap}{\Locks} \step \Conf{T'\upd[e]{t}{\apply{K}{\unit}}}{\Heap'}{\Locks\upd[ls]{\ell}{\holds(t)}}}

  \inferrule*[right=Unlock]
  {T(t) = (\running, e, \Versioned{\delta, c}) \and e \emit{\kw{unlock}\ \ell} K \and \Locks(\ell).ls = \holds(t) \and \Versioned{\Conf{T}{\Heap}{\Locks} \push{t, \ell} \Conf{T'}{\Heap'}{\Locks'}}}
  {\Conf{T}{\Heap}{\Locks} \step \Conf{T'\upd[e]{t}{\apply{K}{\unit}}}{\Heap'}{\Locks'\upd[ls]{\ell}{\kw{free}}}}
\end{mathpar}
}
\caption{Concurrency semantics.}
\Description[Concurrency semantics.]{The figure gives the Start, End, Read, Write, Spawn, Join, Lock, and Unlock rules for the unified concurrent runtime.}
\label{fig:semantics}
\end{subfigure}

\caption{Syntax and semantics. Content in \Versioned{\text{violet}} boxes is only relevant for versioned memory.}
\Description[Syntax and semantics.]{The full figure combines the syntax, sequential effect interface, and concurrent transition rules.}
\end{figure}

The version-controlled memory models are parametric: they share a concurrent runtime, presented in \cref{fig:sequential,fig:semantics}, that abstracts over three auxiliary relations -- commit, push, and pull.
Instantiating these as the identity relation yields a Java-inspired but deliberately simplified shared-memory semantics (\cref{sec:sharedsem}).
We present two reinterpretations that yield \LinearW{} (\cref{sec:linearsem}) and \BranchingW{} (\cref{sec:branchingsem}).
Content in \Versioned{\text{violet}} boxes is not present for shared memory, only the versioned models.

We assume a relation $e \smash{\emit{\alpha}} K$ that steps expression $e$ to continuation $K$, emitting effect $\alpha$.
The effect signals a read, a write, or some synchronisation action that is consumed by the concurrent runtime.
The continuation is a function that consumes a value (including a special unit value $\unit$), an exceptional value or a thread identifier and produces another expression.

A configuration in the concurrent runtime consists of a collection of threads $T$, a heap $\Heap$, a collection of locks $L$ and a map of volatile variables $V$.
A thread has a lifetime state $\mathit{ts}$ which is $\new$ (new), $\running$ (running) or $\terminated$ (terminated), as well as a current expression $e$.
A lock is $\kw{free}$ or is being $\kw{held}$ by some thread.
$L$ initially maps all lock identifiers $l$ to $\kw{free}$.
For the \branching{} model, a lock also stores the last commit pushed to that lock (initially a special root commit $c_{\text{root}}$).
The heap in shared memory is different from that in versioned memory; its definitions are given in \cref{fig:heap-semantics} and explained in their corresponding sections.

\paragraph{Motivation of Scoping}
Our goal is backwards compatibility for race-free executions of programs (\cf{} \cref{thm:equiv}).
According to the JMM, a program is correctly synchronised if and only if all sequentially consistent executions are free of data races.
The JMM further states that if a program has no data races, then all executions of the program will appear to be sequentially consistent~\cite[\S17.4.5]{JMM}.
This motivates picking a simplified, sequentially consistent shared-memory semantics as a baseline to compare our version-controlled memory models to.
For racy executions, we either intentionally break backwards compatibility by throwing an exception or source reads from writes in a manner consistent with the JMM (\cf{} \cref{thm:hbc}).
We also omit volatile variables from the semantics in the paper as their synchronisation actions are already represented through thread and lock operations (\cf{} \cref{tab:syncasshare-java-intuition}).
Volatile variables are included in the mechanised semantics (\cf{} \cref{sec:mechanisation}).

We model conflicting executions as getting stuck.
This keeps the rules concise: a rule guarded by conflict-freedom would otherwise require a complementary conflict case.
We discuss an alternative presentation in \cref{sec:drfexceptions} which models synchronisation conflicts as explicit exceptions.
Importantly, our results transfer to such a presentation as well (this has been verified in Rocq).

\subsection{Shared Memory Concurrency Semantics}
\label{sec:sharedsem}

\cref{fig:semantics}, when ignoring the \Versioned{\text{violet}} boxes, shows the semantics for a concurrent runtime with non-versioned memory.
The step relation $\Conf{T}{\Heap}{\Locks} \step \Conf{T'}{\Heap'}{\Locks'}$ non-deterministically selects a thread $t \in \dom(T)$ and applies
one rule.
We write $T\upd{t}{s}$ for $T$ updated at $t$ with state $s$.
The commit, push and pull relations are all interpreted as the identity relation, meaning for example that $T = T'$, $\Heap = \Heap'$, $L' = L$ and $V' = V$ for $\Conf{T}{\Heap}{\Locks} \smash{\raisebox{-4pt}{$\push{t, \ell}$}} \Conf{T'}{\Heap'}{\Locks'}$.

\cref{fig:shared-heap} defines the shared-memory heap $\Heap$ in the as a map from variables to values with intuitive definitions of reading and writing.
The initial heap $\Heap_0$ is empty.
Reading a variable before it has been initialised is an error, but an underlying semantics with variable scoping could prevent it.

Rules \textsc{Start} and \textsc{End} are internal lifecycle rules: \textsc{Start} transitions a spawned thread from new to running, and \textsc{End} transitions a normally completed running thread to terminated.
We write $T\upd[\mathit{ts}]{t}{\running}$ to update only the $\mathit{ts}$ field of $t$ to $\running$.
The names \textsc{Start} and \textsc{End} denote runtime lifecycle steps, thread creation is represented by \textsc{Spawn}, and observing completion is represented by \textsc{Join}.
Rules \textsc{Read} and \textsc{Write} read and write values from and to the heap.
Note how the abstract sequential semantics emits a read (or write) effect and a continuation $K$.
The continuation is ``filled in'' with the value $v$ from the heap (or $\unit$ for writing) and the resulting expression is reinstated as the expression of the thread.

The \textsc{Spawn} rule creates a new thread with a fresh thread identifier $t'$.
The new thread starts out with the thread state $\new$ (so that rule \textsc{Start} can start it later).
The \textsc{Join} rule lets a thread join some terminated thread.
The continuation $K$ of the joining thread is passed the result value $r$ from the terminated thread.
Rules \textsc{Lock} and \textsc{Unlock} respectively acquire and release a lock.

\begin{figure}
  \begin{subfigure}{1.0\linewidth}
    \begin{mathpar}
      \Heap \in \Var \rightharpoonup \Val \and \Read(\Heap, x) = \Heap(x) \and \Write(\Heap, x, v) = \Heap\upd{x}{v} \and \Heap_0 = \emptyset
    \end{mathpar}
    \caption{Heap semantics in shared memory.}\label{fig:shared-heap}
    \Description[Heap semantics for shared memory.]{The figure presents the definition of the heap as a map from variables to values and the related helper functions. }
  \end{subfigure}

  \begin{subfigure}{1.0\linewidth}
    \begin{mathpar}
      \begin{array}{rcl}
        \mathsf{C} & \in & \mathsf{CommitID} \rightharpoonup (\Var \rightharpoonup \Val)\\
        \mathsf{P} & \subseteq & \mathsf{CommitID} \times \mathsf{CommitID}\\
        \Heap & ::= & (\mathsf{C}, \mathsf{P})\\
        c_1 <_{\Heap} c_2 & \iff & \Heap = (\mathsf{C}, \mathsf{P}) \land (c_1, c_2) \in \mathsf{P}
      \end{array}
      \and
      \begin{array}{l}
        \Heap_0 = (\{c_{\text{root}}\mapsto\emptyset\}, \emptyset)\\\\
        \Write(\Heap, \delta, x, v) = (\Heap, \delta\upd{x}{v})
      \end{array}
      \\
      \text{When}~\Heap = (\mathsf{C}, \mathsf{P}):\\
      \Heap(c) \triangleq \mathsf{C}(c)\and
      \Heap\upd{c}{\delta} \triangleq (\mathsf{C}\upd{c}{\delta}, \mathsf{P})\and
      \Heap\anc{c}{c'} \triangleq (\mathsf{C}, \mathsf{P} \cup \{(c, c')\})\\
      \Read(\Heap, \delta, c, x) =
      \begin{cases}
        \delta(x) & \text{if $x\in\dom(\delta)$}\\
        v' & \text{if $\Heap(c')(x) = v'$ and $c' \in \max_\Heap(\{c'' \mid x \in \dom(\Heap(c'')) \land c'' <_{\Heap}^* c\})$}
      \end{cases}
    \end{mathpar}
    \caption{Heap semantics in versioned memory. }\label{fig:versioned-heap}
    \Description[Heap semantics in versioned memory.]{The figure presents the definition of the heap as a graph and the related helper functions. }
  \end{subfigure}
  \caption{Heap semantics and related helper functions.}\label{fig:heap-semantics}
  \Description[Heap semantics and related helper functions.]{The full figure presents the heap semantics for shared memory and the versioned memory. }
\end{figure}

\subsection{Version-Controlled Heap Semantics}
\label{sec:graph-semantics}

In versioned memory, the heap is a graph of commits where each commit is mapped to a diff.
\cref{fig:versioned-heap} shows the definition of this graph and the related helper functions.
Given a commit graph $\Heap$, we define an ancestor relation $\cdot <_{\Heap} \cdot$ for when a commit is the direct ancestor of another (we use $<_{\Heap}^*$ for the reflexive and transitive closure).
The initial graph $\Heap_0$ contains a single root commit with an empty diff (again, reading an uninitialised variable is an error).

Each thread has a local staging area $\delta$ with writes that have not yet been committed to memory, and a commit $c$ which is the latest commit that thread has observed.
All writes of a thread go to its staging area, which is later committed to the heap during synchronisation.
When a thread reads a value, it checks its own staging area first and then finds the latest value in the heap according to the latest commit the thread has observed.
The function $\max_\Heap$, used in the \Read{} helper, is defined as the maximal element of a set of commits with respect to $<_{\Heap}^*$.
While this partial order allows multiple maximal elements -- this would be the result of an execution with conflicting writes -- both version-controlled models guarantee that, unless a conflict is detected, the set of latest values is never larger than one element (\cf{} \cref{lem:read_inj}).

In order to detect conflicts in the commit graph, we define the change set $\changes{\Heap}{c_1} c_2$ as the set of variables modified in the history of $c_2$ but not visible in the history of $c_1$:
\begin{definition}[Change set]
  $
    \changes{\Heap}{c_1} c_2 =
    \{ x \mid \exists c. c <_{\Heap}^* c_2 \land \lnot(c <_{\Heap}^* c_1) \land x \in \dom(\Heap(c)) \}
  $
\end{definition}
What sets the two version-controlled models apart is the shape of the commit graphs, their implementation of synchronisation actions, and when they check for conflict.

\subsection{Linear Memory Concurrency Semantics}
\label{sec:linearsem}

In the \linear{} model, the commit graph is linear, with the latest commit always being the last commit made by any thread.
Whenever a thread synchronises with the heap, the concurrent runtime checks the staging area for conflicts (we write $c_{\max}(\Heap) $ for the maximal commit in $\Heap$).
\begin{definition}[Conflict-free commit]
  $
    \text{Conflict-Free}(\delta, c, \Heap) \iff \dom(\delta) \cap \changes{\Heap}{c}{(c_{\max}(\Heap))} = \emptyset
  $
\end{definition}
On thread initialisation, starting threads get the maximal heap commit
\(
  \Init(H, c) = c_{\max}(\Heap).
\)

A thread can pull in the updates from the heap by checking its staging area for conflicts and fast-forwarding its own commit to the current global maximum (the synchronisation object is irrelevant in \LinearW{} and is ignored).
\begin{mathpar}
\inferrule*[right=L-Pull]
{T(t) = (\mathit{ts}, e, \delta, c) \and \text{Conflict-Free}(\delta, c, \Heap)}
{\Conf{T}{\Heap}{\Locks} \pull{t, \_} \Conf{T\upd[c]{t}{c_{\max}(\Heap)}}{\Heap}{\Locks}}
\end{mathpar}

When a thread commits its staged updates it moves its staging area to the heap as a fresh commit with the current global maximal commit as its ancestor.
\begin{mathpar}
\inferrule*[right=L-Commit]
{T(t) = (\mathit{ts}, e, \delta, c) \and \text{Conflict-Free}(\delta, c, \Heap) \and c' \notin \dom(\Heap)}
{\Conf{T}{\Heap}{\Locks} \commit{t} \Conf{T\upd{t}{(\mathit{ts}, e, \emptyset, c')}}{\Heap\upd{c'}{\delta}\anc{c_{\max}(\Heap)}{c'}}{\Locks}}
\end{mathpar}

When a thread synchronises with another thread or lock it commits its staged updates (ignoring the synchronisation object $t'/\ell$).
\begin{mathpar}
\inferrule*[right=L-PullPush]
{\Conf{T}{\Heap}{\Locks} \commit{t} \Conf{T'}{\Heap'}{\Locks}}
{\Conf{T}{\Heap}{\Locks} \push{t,\_} \Conf{T'}{\Heap'}{\Locks}}
\end{mathpar}

With these auxiliary relations in place, the rules in \cref{fig:semantics} define how threads synchronise across different actions.
The \textsc{Start} rule starts a thread and sets its commit to the current maximum.
The \textsc{End} rule terminates a thread and also commits its updates.
The \textsc{Spawn} rule commits the updates of the spawning thread via \textsc{L-PullPush} and spawns a new thread that starts with an empty staging area and a dummy commit.
The dummy commit will be updated by the \textsc{Start} rule, and no other rule can observe a thread with the thread state $\new$, so this is safe.
The \textsc{Join} rule uses \textsc{L-Pull} to pull from the heap, which contains all updates of the joined thread due to \textsc{End}.
The two \textsc{Lock} rules pull the latest commit via \textsc{L-Pull} and the \textsc{Unlock} rule commits via \textsc{L-PullPush}.

\subsection{Branching Memory Concurrency Semantics}
\label{sec:branchingsem}

In the \branching{} model the commit graph is a DAG.
Committing in the \branching{} model commits the staging area of a thread as a child of the thread's current commit and updates it.
It does not perform a conflict check because the updates are not yet visible to other threads.
\begin{mathpar}
\inferrule*[right=B-Commit]
{
  T(t) = (\running, e, \delta, c) \\
  c_{\text{new}} \not\in \dom(\Heap) \\
  \Heap' = \Heap\upd{c_{\text{new}}}{\delta}\anc{c}{c_{\text{new}}}
}
{\Conf{T}{\Heap}{\Locks} \commit{t} \Conf{T\upd{t}{(\running, e, \emptyset, c_{\text{new}})}}{\Heap'}{\Locks}}
\end{mathpar}

In the \branching{} model, both threads and locks have a current commit.
We define two actions for committing a thread's staged updates to the commit graph and communicating this commit to another thread or lock.
\begin{mathpar}
\inferrule*[right=B-Push-Thread]
{
  T(t) = (\running, e, \delta, c) \\
  \Conf{T}{\Heap}{\Locks} \commit{t} \Conf{T'}{\Heap'}{\Locks} \\
  t'\in\dom(T)
}
{\Conf{T}{\Heap}{\Locks} \push{t, t'} \Conf{T'\upd[c]{t'}{T'(t).c}}{\Heap'}{\Locks}}

\inferrule*[right=B-Push-Lock]
{
  T(t) = (\running, e, \delta, c) \\
  \Conf{T}{\Heap}{\Locks} \commit{t} \Conf{T'}{\Heap'}{\Locks} \\
  \ell\in\dom(\Locks)
}
{\Conf{T}{\Heap}{\Locks} \push{t, \ell} \Conf{T'}{\Heap'}{\Locks \upd[c]{\ell}{T'(t).c}}}

\end{mathpar}

Similarly, a thread can pull from another thread or from a lock.
In both cases, the runtime checks for conflicts before creating a merge commit.
We define the condition under which a thread with staging area $\delta$ and commit $c_1$ can merge with another commit $c_2$ without conflicts:
\begin{definition}[Conflict-free merge of two commits]
  \[
    \text{Conflict-Free-Merge}(\delta, c_1, c_2, \Heap) \iff (\dom(\delta) \cup \changes{\Heap}{c_2} c_1) \cap \changes{\Heap}{c_1} c_2 = \emptyset
  \]
\end{definition}
This requires that the updates introduced by $c_2$ do not conflict with the thread's current commit $c_1$, nor with the uncommitted writes in its staging area $\delta$.
We define an auxiliary merge operation that checks for conflicts and creates a merge commit.
\begin{mathpar}
\inferrule*[right=B-Merge]
{
  T(t) = (\running, e, \delta, c_t) \\
  \text{Conflict-Free-Merge}(\delta, c_t, c, \Heap) \\
  c_{\text{new}} \not\in \dom(\Heap) \\
  \Heap' = \Heap\upd{c_{\text{new}}}{\emptyset}\anc{c_t}{c_{\text{new}}}\anc{c}{c_{\text{new}}}\\
}
{
  \Conf{T}{\Heap}{\Locks} \merge{t}{c} \Conf{T\upd{t}{(\running, e, \delta, c_{\text{new}})}}{\Heap'}{\Locks}
}
\end{mathpar}

The pull operation merges with a specified external commit.
Unlike \LinearW{}, which implicitly refers to the global history, pulling in \BranchingW{} requires an explicit target.
\begin{mathpar}
\inferrule*[right=B-Pull-Thread]
{
  T(t') = (\_,\_,\_,c') \and
  \Conf{T}{\Heap}{\Locks} \merge{t}{c'} \Conf{T'}{\Heap'}{\Locks}
}
{
  \Conf{T}{\Heap}{\Locks} \pull{t, t'}  \Conf{T'}{\Heap'}{\Locks}
}

\inferrule*[right=B-Pull-Lock]
{
  \Locks(\ell) = (\_, c) \and
  \Conf{T}{\Heap}{\Locks} \merge{t}{c} \Conf{T'}{\Heap'}{\Locks}
}
{
  \Conf{T}{\Heap}{\Locks} \pull{t, \ell} \Conf{T'}{\Heap'}{\Locks}
}

\end{mathpar}
Pulling leaves staged updates uncommitted, mirroring the pull action in the linear model.

When starting, a thread keeps the commit given to it by its spawning thread: \(\Init(H, c) = c.\)

The synchronisation in \cref{fig:semantics} is the same as for the \linear{} model; the differences are in the definitions of the synchronisation actions.
The \textsc{Spawn} rule pushes the thread's staged updates to the commit graph and spawns a new thread.
The new thread is spawned with the same commit as the parent thread (via \textsc{B-Push-Thread}), so it inherits the parent's private view.
This reflects a branching point in the version history, where the new thread can diverge from the parent thread's view and make independent updates.
The \textsc{Join} rule pulls the last commit from the joined thread into the joining thread before retrieving the result of the joined thread.
This is a merge point between two divergent branches in the commit graph, where the joining thread synchronises with the updates made by the joined thread.
The \textsc{Lock} rules atomically pulls the latest commit from the lock's commit into the locking thread and acquires the lock.
This ensures that the locking thread synchronises with any updates that have been pushed to the lock before acquiring it.
The \textsc{Unlock} rule pushes the unlocking thread's staged updates to the commit graph and releases the lock.
The rule updates the lock's commit to the unlocking thread's new commit, so updates made while holding the lock are available to the next acquirer.

\subsubsection*{A Worked Example}
\label{sec:workedexample}
To build intuition for the branching semantics, we revisit the nested-locking example from the end of \cref{sec:syncasshare}.
\cref{fig:nested-locking-branching} shows the program code and the corresponding updates to the commit graph.
Acquiring lock $l_1$ pulls from $l_1$ by creating a merge commit from $T$'s current commit and $l_1$'s current commit, and advances $T$ to that merge commit.
The write to $x$ is added to $T$'s staging area.
Acquiring $l_2$ proceeds analogously, creating a merge commit from $T$'s current commit and $l_2$'s current commit.
The write to $y$ is also added to $T$'s staging area.
Unlocking $l_2$ commits $T$'s staged writes, $\{x,y\}$, as a child of $T$'s current commit and advances $l_2$ to that new commit.
After the write to $z$, unlocking $l_1$ commits the staged write $\{z\}$ as a descendant of the commit published to $l_2$, and advances $l_1$ to it.
This is precisely the release-acquire guarantee: a thread acquiring $l_2$ sees all writes committed during the unlock of $l_2$, including the $x$ written before $l_2$ was even locked.
Information transfer through synchronisation is naturally realised in version-controlled memory: when a thread pushes to a lock, it publishes the thread's current commit, including writes performed before that lock was acquired.

\begin{figure}[t]
  \centering
  \small
  \begin{tabular}{p{0.28\linewidth}@{\hspace{0.5em}}p{0.16\linewidth}@{\hspace{0.5em}}p{0.48\linewidth}}
    \toprule
    Code & Action & Commit graph \\
    \midrule
    --- & --- & initial: $T=\Commit{1}$, $l_1=\Commit{2}$, $l_2=\Commit{3}$ \\
    \lstinline[style=SmallCode,language=Java]|synchronized (l1) {| & lock $l_1$ & $T$ creates $\Commit{4}$, child of $\Commit{1}$ and $\Commit{2}$; advances to $\Commit{4}$ \\
    \hspace{1.5ex}\lstinline[style=SmallCode,language=Java]|x = 1;| & write $x$ & $T$ stages $x$ \\
    \hspace{1.5ex}\lstinline[style=SmallCode,language=Java]|synchronized (l2) {| & lock $l_2$ & $T$ creates $\Commit{5}$, child of $\Commit{4}$ and $\Commit{3}$; advances to $\Commit{5}$ \\
    \hspace{3ex}\lstinline[style=SmallCode,language=Java]|y = 2;| & write $y$ & $T$ stages $y$ \\
    \hspace{1.5ex}\lstinline[style=SmallCode,language=Java]|}| & unlock $l_2$ & $T$ commits $\Commit{6}$ with $\{x,y\}$; $l_2$ advances to $\Commit{6}$ \\
    \hspace{1.5ex}\lstinline[style=SmallCode,language=Java]|z = 3;| & write $z$ & $T$ stages $z$ \\
    \lstinline[style=SmallCode,language=Java]|}| & unlock $l_1$ & $T$ commits $\Commit{7}$ with $\{z\}$; $l_1$ advances to $\Commit{7}$ \\
    \bottomrule
  \end{tabular}
  \caption{Version-controlled representation of nested locking in \BranchingW{}. A single thread $T$ acquires two locks, $l_1$ and $l_2$; each lock receives the thread's current commit, including writes performed before that lock was acquired.}
  \Description[Nested locking in Branching version-controlled memory.]{A table shows a single thread locking l1, writing x, locking l2, writing y, unlocking l2, writing z, and unlocking l1. Unlocking l2 publishes a commit containing x and y; unlocking l1 later publishes a descendant commit containing x, y, and z.}
  \label{fig:nested-locking-branching}
\end{figure}

The commit graph preserves per-synchronisation-object history without discarding the thread-local program-order history.
The same reasoning extends to hand-over-hand locking: successive unlocks advance the corresponding locks to the commits current at those releases.

\subsection{Detecting and Reporting Conflicts}
\label{sec:drfexceptions}

We model conflicting executions as getting stuck.
An alternative presentation could include failing variants of every conflict-checking rule that provides an exceptional value $\mathit{exc}$ to the continuation $K$.
This would be closer to a user-facing implementation of version-controlled memory.
For example, the rule for a conflicting join in the branching semantics could be written as follows:
\begin{mathpar}
  \inferrule*[right=JoinConflict]
  {T(t) = (\running, e, \delta, c) \and e \emit{\kw{join}\ t'} K \and
    T(t') = (\terminated, r, \_, c') \and\\\\
    \lnot\text{Conflict-Free-Merge}(\delta, c, c', \Heap)}
  {\Conf{T}{\Heap}{\Locks} \step \Conf{T\upd[e]{t}{\apply{K}{\mathit{exc}}}}{\Heap}{\Locks}}
\end{mathpar}
Our proof of backwards-compatibility transfer to such a presentation as we only care about race-free executions, where a merge will never fail (\cf{} \cref{lem:cf-merge}).
Whether the failing thread should terminate and lose its staged updates or recover from a detected conflict (catching the exception) is an open question that we leave for future work.

\section{Metatheory}
\label{sec:metatheory}

In this section we discuss the metatheoretical properties of the formalism in \cref{sec:formal-semantics}.
Our main result is that for race-free executions both version-controlled memory models are equivalent to the shared-memory model (\cref{thm:equiv}).
We also show that threads that read initialised variables always read values permitted by the JMM, even in the presence of races (\cref{thm:hbc}).
To formulate these properties, we first define a variant of each of the three semantics that emits a log of events in the order they happened (\cref{sec:logging}).
We then define what it means for a log to be race free and show that all three semantics coincide for executions that produce a race-free log (\cref{sec:racefree}).
All definitions and proofs have been mechanised in Rocq~\cite{vcmRocq}, including volatile variables, which are omitted from the paper presentation (\cref{sec:mechanisation}).

\subsection{Event Logs and Logging Semantics}
\label{sec:logging}

\begin{figure}[t]
\begin{mathpar}
  \begin{array}{rcl}
    E & ::= & t~\Start ~|~ t~\End ~|~ t~\Write~x~v ~|~ t~\Read~x~v ~|~ t~\Spawn~t ~|~ t~\EJoin~t ~|~ t~\ELock~\ell ~|~ t~\Unlock~\ell\\
    \Log & ::= & \Log, E ~|~ \epsilon\\
  \end{array}

  \inferrule*[right=HB-PO]
  {
    \Log[m] = E_m \and \Log[n] = E_n\\\\
    E_m.t = E_n.t \and m < n
  }
  {
    \Log \vdash m \prec n
  }
  \quad
  \inferrule*[right=HB-Lock]
  {
    \Log[m] = \_~\Unlock~\ell \and \Log[n] = \_~\ELock~\ell\\\\
    m < n
  }
  {
    \Log \vdash m \prec n
  }

  \inferrule*[right=HB-Spawn]
  {
    \Log[m] = \_~\Spawn~t \and \Log[n] = t~\Start\\\\
    m < n
  }
  {
    \Log \vdash m \prec n
  }
  \quad
  \inferrule*[right=HB-Join]
  {
    \Log[m] = \End~t \and \Log[n] = \_~\EJoin~t\\\\
    m < n
  }
  {
    \Log \vdash m \prec n
  }

  \inferrule*[right=HB-Trans]
  {
    \Log \vdash m \prec k \and
    \Log \vdash k \prec n
  }
  {
    \Log \vdash m \prec n
  }
\end{mathpar}

  \caption{Syntax of events and definition of happens-before relation on logs.}
  \Description[Syntax of events and definition of happens-before relation on logs.]{The figure presents the syntax of events and the five derivation rules for the happens-before relation on logs. }
  \label{fig:log}
\end{figure}

To reason about whether a data race has occurred, it is not enough to look for two threads about to access the same memory.
Any number of operations could have happened between two racing accesses.
Instead, the semantics produces a log of events, which we inspect to decide whether an unsynchronised access has occurred.
\cref{fig:log} shows the syntax of logs.
There is one event per rule in the semantics, and each event contains the thread that caused the event.

We omit the well-formedness relation for logs, written $\vdash \Log~\mathbf{wf}$.
It requires unique spawn and start events; threads that start only after spawning and have no events before their start or after their end; joins only after the joined thread has ended; and locks acquired and released in order by the same thread.
To avoid the chicken-and-egg problem of requiring every thread to be spawned before it starts, we require a non-empty log to begin with the start event of a special root thread $t_{\text{root}}$, which was never spawned.
See the mechanisation for details (\cref{sec:mechanisation}).

\cref{fig:log} also shows a happens-before relation on events in a log, defining when two events are definitely ordered.
Because the same event can occur multiple times in a log, it is based on the indices of the events.
$\Log \vdash m \prec n$ should be read as ``in $\Log$, the event at index $m$ happens before the event at index $n$''
Rule \textsc{HB-PO} orders two events performed by the same thread by program order.
Rule \textsc{HB-Lock} orders each unlock event before subsequent lock events of the same lock.
Rules \textsc{HB-Spawn} and \textsc{HB-Join} orders the spawn event of a thread before its start event, and the end of a thread before subsequent join events.
Rule \textsc{HB-Trans} makes the happens-before relation transitive.

For each relation $\mathit{cfg} \step \mathit{cfg}'$ we define the relation $\mathit{cfg} ~|~ \Log \step \mathit{cfg}' ~|~ \Log'$ that is equivalent but additionally emits the corresponding event to its log.
Proving this equivalence is straightforward.

\begin{theorem}[Equivalence of logging semantics]
  The logging semantics is equivalent to the non-logging semantics (for all three models)
  \[
    \mathit{cfg} \step \mathit{cfg}' \iff \exists \Log'. \mathit{cfg} ~|~ \Log \step \mathit{cfg}' ~|~ \Log'
  \]
\end{theorem}

All three models have a relation $\Log \vdash \mathit{cfg}$ that defines when a log correctly describes a configuration.
We omit the formal definition of the shared parts here:
threads with a spawn event have spawned,
spawned threads without a start event have the new lifetime state ($\new$),
threads with an end event have terminated (have lifetime state $\terminated$),
threads that have a start event but no end event are running (have lifetime state $\running$),
locks whose latest event was a lock are locked,
and locks without events or whose latest event was an unlock are free.

For the logging shared-memory semantics, $\Log \vdash \mathit{cfg}$ also requires that the heap maps every variable to the value given by the last write event touching that variable.

\begin{definition}[The log describes the shared-memory heap]\label{def:last-write}
  \[
    \begin{array}{l}
      \Log \vdash \Conf{T}{H}{L} \iff ...~\land\\
      \quad\left(H(x) = v \iff \exists \Log'.~[E \mid E\in\Log \land E = \_~\Write~x~\_] = \Log', \_~\Write~x~v\right)
    \end{array}
  \]
\end{definition}

The two versioned semantics need more structure, since every thread should either see its own latest value or the latest global value seen from the thread's commit.
To capture this, $\Log \vdash \mathit{cfg}$ for the two versioned semantics includes the requirement that replaying the log reproduces a $\View$ that is consistent with $\mathit{cfg}$.
(We focus on the branching model here; the definitions for the linear model are similar. )

\begin{figure}

\begin{mathpar}
  \Delta \in \TID \rightharpoonup (\Var \rightharpoonup \Val)\and
  TC \in \TID \rightharpoonup \CommitID\and
  LC \in \Lock \rightharpoonup \CommitID\\
  CH ::= CH, c ~|~ \epsilon \and \View ::= \langle\Delta, TC, LC, \Heap\rangle\\
  \inferrule*[right=Play-Start]
  {
  }
  {
    t~\Start ~/~ \langle \Delta; TC; LC; \Heap\rangle \Play \langle \Delta; TC; LC; \Heap\rangle
  }

  \inferrule*[right=Play-End]
  {
    c' \not\in \dom(\Heap) \\\\
    \Heap' = \Heap\upd{c'}{\Delta(t)}\anc{TC(t)}{c'}
  }
  {
    t~\End ~/~ \langle \Delta; TC; LC; \Heap\rangle \Play \langle \Delta\upd{t}{\emptyset}; TC\upd{t}{c'}; LC; \Heap'\rangle
  }\\

  \inferrule*[right=Play-Read]
  {
  }
  {
    t~\Read~x~v ~/~ \langle \Delta; TC; LC; \Heap\rangle \Play \langle \Delta; TC; LC; \Heap\rangle
  }

  \inferrule*[right=Play-Write]
  {
    \Delta' = \Delta\upd{t}{\Delta(t)\upd{x}{v}}
  }
  {
    t~\Write~x~v ~/~ \langle \Delta; TC; LC; \Heap\rangle \Play \langle \Delta'; TC; LC; \Heap\rangle
  }\\

  \inferrule*[right=Play-Spawn]
  {
    c' \not\in \dom(C) \\\\
    \Delta' = \Delta\upd{t_1}{\emptyset}\upd{t_2}{\emptyset} \\\\
    TC' = TC\upd{t_1}{c'}\upd{t_2}{c'}\\\\
    \Heap' = \Heap\upd{c'}{\Delta(t_1)}\anc{TC(t)}{c'}
  }
  {
    t_1~\Spawn~t_2 ~/~ \langle \Delta; TC; LC; \Heap\rangle \Play \langle \Delta'; TC'; LC; \Heap'\rangle
  }

  \inferrule*[right=Play-Join]
  {
    c' \not\in \dom(\Heap) \\\\
    \Heap' = \Heap\upd{c'}{\emptyset}\anc{TC(t_1)}{c'}\anc{TC(t_2)}{c'}
  }
  {
    t_1~\EJoin~t_2 ~/~ \langle \Delta; TC; LC; \Heap\rangle \Play \langle \Delta; TC\upd{t_1}{c'}; LC; \Heap'\rangle
  }

  \inferrule*[right=Play-LockFresh]
  {
    l \not\in \dom(LC)
  }
  {
    t~\ELock~\ell ~/~ \langle \Delta; TC; LC; \Heap\rangle \Play \langle \Delta; TC; LC\upd{l}{TC(l)}; \Heap\rangle
  }

  \inferrule*[right=Play-Lock]
  {
    l \in\dom(LC) \and c' \not\in \dom(\Heap) \\\\
    \Heap' = \Heap\upd{c'}{\emptyset}\anc{TC(t)}{c'}\anc{LC(\ell)}{c'}
  }
  {
    t~\ELock~\ell ~/~ \langle \Delta; TC; LC; \Heap\rangle \Play \langle \Delta; TC; LC\upd{l}{c'}; \Heap'\rangle
  }

  \inferrule*[right=Play-Unlock]
  {
    c' \not\in \dom(C) \\\\
    \Heap' = \Heap\upd{c'}{\Delta(t)}\anc{TC(c)}{c'}
  }
  {
    t~\Unlock~\ell ~/~ \langle \Delta; TC; LC; \Heap\rangle \Play \langle \Delta\upd{t}{\emptyset}; TC\upd{t}{c'}; LC\upd{\ell}{c'}; \Heap'\rangle
  }

  \inferrule*[right=Play-Snoc]
  {
    \Log ~/~ \langle \Delta; TC; LC; \Heap\rangle \Play CH ~/~ \langle \Delta'; TC'; LC'; \Heap'\rangle \\\\
    E ~/~ \langle \Delta'; TC'; LC'; \Heap'\rangle \Play \langle \Delta''; TC''; LC''; \Heap''\rangle
  }
  {
    \Log, E ~/~ \langle \Delta; TC; LC; \Heap\rangle \Play CH, TC''(E.t) ~/~  \langle \Delta''; TC''; LC''; \Heap''\rangle
  }

  \inferrule*[right=Play-Nil]
  {
  }
  {
    \epsilon ~/~ \langle \Delta; TC; LC; \Heap\rangle \Play \epsilon ~/~ \langle \Delta; TC; LC; \Heap\rangle
  }\\
\end{mathpar}

  \caption{Rules for replaying a log for the branching semantics (\linear{} is similar).}
  \Description[Rules for replaying a log for the branching semantics.]{The figure presents four rules for replaying a single event, and the two rules for replaying a log. It also shows the related part of when a log correctly describes a configuration. }
  \label{fig:play}
\end{figure}

A $\View$ is a tuple $\langle\Delta, TC, LC, H\rangle$ where $\Delta$ maps thread identifiers to their uncommitted staging areas, $TC$ maps thread identifiers to their commits, $LC$ maps locks to their commits, and $H$ is the commit graph from \cref{sec:graph-semantics}.
\cref{fig:play} shows the replay relation for single events, written as $E ~/~ \View \rhd \View'$.
Each rule of the replay relation mirrors a rule in \cref{fig:semantics}, capturing the changes to staging areas and commits.
To exemplify the rules in \cref{fig:play} we describe four rules here.
Rules \textsc{Play-Start} and \textsc{Play-Read} do not update any commit information.
Rule \textsc{Play-End} commits the staged updates $\Delta(t)$ of a terminating thread $t$ as a fresh child commit $c'$ of $t$'s current commit $TC(t)$ and updates the staging area and commit of $t$ in $\Delta$ and $TC$.
Rule \textsc{Play-Write} updates the staging area of $\Delta(t)$ so that it also includes the updated variable.
Rules \textsc{Play-Snoc} and \textsc{Play-Nil} lift the other replay rules to a log, $\Log ~/~ \View \rhd CH ~/~ \View'$, additionally recording the commit of each event in the log in $CH$ (this is used for bookkeeping commit ancestry in the proofs).
For example, when the event $\End~t$ is replayed, $CH$ will be extended with the fresh commit given to the terminating thread.

The relation $\Log \vdash \mathit{cfg}$ requires that replaying the log from an initial view $\View_0$ produces a $\View$ that is consistent with the configuration: in addition to producing the same graph, each thread holds a staging area and commit that agrees with $\Delta$ and $TC$, and each lock holds a commit that agrees with $LC$.
The initial view $\View_0$ maps the root thread to an empty staging area, the root thread and all locks to the root commit $c_{\text{root}}$, and contains the initial commit graph $\Heap_0$.

\begin{definition}[Replaying the log creates a consistent view]\label{def:replay}

  \begin{mathpar}
    \begin{array}{l}
      \Log \vdash \Conf{T}{H}{L} \iff ... ~\land~\\\\\\\\
    \end{array}
    \begin{array}{l}
      \exists \Delta, TC, LC, CH.\\
      \quad \Log ~/~ \View_0 \Play CH ~/~  \langle \Delta; TC; LC; \Heap\rangle ~\land\\
      \quad (\forall t \in\dom(T) . \exists \mathit{ts}, e.~ T(t) = (\mathit{ts}, e, \Delta(t), TC(t))) ~\land\\
      \quad (\forall \ell \in\dom(L) . \exists \mathit{ls}.~ L(\ell) = (\mathit{ls}, LC(\ell)))
    \end{array}\\
    \View_0 = \langle \{t_{\text{root}}\mapsto \emptyset\}; \{t_{\text{root}}\mapsto c_{\text{root}}\}; LC_0; \Heap_0\rangle\and LC_0(\ell) = c_{\text{root}}
  \end{mathpar}
\end{definition}

For all three logging semantics, we prove that the dynamic semantics produce well-formed logs describing the resulting configurations.

\begin{theorem}[Well-formedness of initial configuration]
  The initial configuration, containing an arbitrary expression $e$, is described by the empty log (for all three models)
  \begin{mathpar}
    \epsilon \vdash \mathit{cfg}_0 \quad \text{where}~
    \mathit{cfg}_0 = \Conf{\{t_{\text{root}} \mapsto (\new, e, \Versioned{\emptyset, c_{\text{root}}})\}}{\Heap_0}{L_0}~
    \text{and}~L_0(\ell) = (\kw{free}, \Versioned{c_{\text{root}}})
  \end{mathpar}
\end{theorem}

\begin{theorem}[Preservation of logging semantics]\label{thm:preservation}
  The logging semantics preserves well-formed logs that describe configurations (for all three models)
  \[
    \vdash \Log~\mathbf{wf} ~\land~ \Log \vdash \mathit{cfg} ~\land~ \mathit{cfg} ~|~ \Log \step \mathit{cfg}' ~|~\Log'  \implies \vdash \Log'~\mathbf{wf} ~\land~ \Log' \vdash \mathit{cfg}'
  \]
\end{theorem}

Before turning to race-free executions, note that our versioned semantics is also backwards compatible in the presence of conflicts that have not yet caused an exception.
The JMM requires that the result of a read $r$ is sourced from a write $w$ such that $r$ does not happen-before $w$ and such that there is no write $w'$ such that $w$ happens-before $w'$ and $w'$ happens-before $r$~\cite[\S17.4.5]{JMM}.
We define this property for logs and prove that it is preserved by our two versioned semantics, even in the presence of races.
\begin{definition}[Happens-before consistency]
  \[
    \begin{array}{l}
      \vdash \Log~\mathbf{hbc} \iff \forall n, x, v.\\
      \quad \Log[n] = \_~\Read~x~v ~\implies\\
      \qquad \exists m.~ \Log[m] = \_~\Write~x~v ~\land~
      \lnot (\Log \vdash n \prec m) ~\land~\\
      \qquad\quad\lnot (\exists k. Log[k] = \_~\Write~x~\_ ~\land~ \Log \vdash m \prec k ~\land~ \Log \vdash k \prec n)\\
    \end{array}
  \]
\end{definition}
\begin{theorem}[Preservation of happens-before consistency]\label{thm:hbc}
  If a well-formed and happens-before consistent log describes a configuration and that configuration can take a step, then the resulting log is also happens-before consistent.
  \[
    \begin{array}{l}
      \vdash \Log~\mathbf{wf} ~\land~
      \Log \vdash \mathit{cfg} ~\land~
      \vdash \Log~\mathbf{hbc} ~\land~
      \mathit{cfg} ~|~ \Log \step \mathit{cfg}' ~|~ \Log' ~\implies\\
      \quad \vdash \Log~\mathbf{hbc}
    \end{array}
  \]
\end{theorem}

\subsection{Race-Free Executions}
\label{sec:racefree}

A log describing the history of a program, together with a definition of when events are ordered (\cref{fig:log}), lets us define what it means for an execution to be race free.

\begin{definition}[Race-free log]
  A log is race-free, written $\vdash \Log~\mathbf{rf}$, if for every pair of a write event and a read or write event of the same variable the first event happens before the second.
  \[
    \begin{array}{l}
      log[m] = \_~\Write~x~\_ ~\land~
      (log[n] = \_~\Read~x~\_ ~\lor~ log[n] = \_~\Write~x~\_) ~\land~
      m < n \implies\\
      \quad \Log \vdash m \prec n
    \end{array}
  \]
\end{definition}

A race-free log describes a race-free execution, where reads and writes to the same variable are ordered by happens-before.
This excludes, for example, the execution in \cref{fig:linear-counter-conflict} where the two writes in the different threads are unordered (so is the write by thread 1 and the read by thread 2).

Replaying a race-free log in the versioned semantics produces configurations with several important properties.
First, the value of a variable in versioned memory is the same as the last write to that variable.

\begin{lemma}[Last versioned value is last write in race-free executions]\label{lem:read}
  When replaying a well-formed, race-free log that ends with thread $t$ reading $v$ from variable $x$, looking up the latest value of $x$ gives the same value as the last write to that variable.
  \[
    \begin{array}{l}
      \vdash \Log, t~\Read~x~v~\mathbf{wf}~\land~
      \vdash \Log, t~\Read~x~v~\mathbf{rf}~\land~
      \Log ~/~ \View_0 \Play CH ~/~  \langle \Delta; TC; LC; \Heap\rangle~\implies\\
      \quad\Read(\Heap, \Delta(t), TC(t), x) = v \iff
      \exists \Log'.~[E \mid E\in\Log \land E = \_~\Write~x~\_] = \Log', \_~\Write~x~v
    \end{array}
  \]
\end{lemma}

The lemma contains two cases, based on whether the $\Read$ goes via the local staging area $\Delta(t)$ or the commit graph $H$.
Having the read event in the log is essential to guarantee that the write to $x$ happens before the read.
Note that the right-hand side of the equivalence in the conclusion of this lemma matches the right-hand side of the equivalence in \cref{def:last-write}.
This lets us relate the versioned-memory heap to the shared-memory heap.

A related lemma states that reads are deterministic for race-free executions.
\begin{lemma}[Deterministic reads in race-free executions]\label{lem:read_inj}
  When replaying a well-formed, race-free log, looking up the latest value of a variable always gives the same value.
  \[
    \begin{array}{l}
      \vdash \Log~\mathbf{wf}~\land~
      \vdash \Log~\mathbf{rf}~\land~
      \Log ~/~ \View_0 \Play CH ~/~  \langle \Delta; TC; LC; \Heap\rangle~\land\\
      \Read(\Heap, \delta, c, x) = v_1~\land~
      \Read(\Heap, \delta, c, x) = v_2~\implies\\
      \quad v_1 = v_2
    \end{array}
  \]
\end{lemma}

Furthermore, committing (in the linear model) and merging (in the branching model) always succeeds in a race-free execution.

\begin{lemma}[Committing in the linear model for race-free executions]\label{lem:cf-commit}
  When replaying a well-formed, race-free log, any thread can commit its staging area as the new global maximum commit without conflict.
  \[
    \begin{array}{l}
      \vdash \Log~\mathbf{wf}~\land~
      \vdash \Log~\mathbf{rf}~\land~
      \Log ~/~ \View_0 \Play CH ~/~  \langle \Delta; TC; LC; \Heap\rangle~\implies\\
      \quad \text{Conflict-Free}(\Delta(t), TC(t), \Heap)
    \end{array}
  \]
\end{lemma}

\begin{lemma}[Merging in the branching model for race-free executions]\label{lem:cf-merge}
  When replaying a well-formed, race-free log, any thread can merge its staging area with any commit in the commit graph without conflict.
  \[
    \begin{array}{l}
      \vdash \Log~\mathbf{wf}~\land~
      \vdash \Log~\mathbf{rf}~\land~
      \Log ~/~ \View_0 \Play CH ~/~  \langle \Delta; TC; LC; \Heap\rangle~\land
      c\in\dom(H)\implies\\
      \quad \text{Conflict-Free-Merge}(\Delta(t), TC(t), c, \Heap)
    \end{array}
  \]
\end{lemma}

The contrapositives of these two lemmas are also informative: if a conflict check fails, there must have been an actual data race.
We show this here for joining a thread.
\begin{corollary}[Soundness of conflict detection for joins]
  If trying to join a terminated thread fails due to a reported conflict, there was an actual data race.
  \[
    \begin{array}{l}
      \vdash \Log~\mathbf{wf}~\land~
      \Log \vdash \Conf{T}{H}{L}~\land~
      T(t) = (\running, e, \delta, c)~\land~
      e \emit{\kw{join}\ t'} K~\land\\
      T(t') = (\terminated, r, \_, c')~\land~
      \lnot\text{Conflict-Free-Merge}(\delta, c, c', \Heap)\implies\\
      \quad \lnot (\vdash \Log~\mathbf{rf})
    \end{array}
  \]
\end{corollary}
\noindent
Similar lemmas can be stated about all rules that contain a conflict check.

Our main theorem, which is identical for both the linear and branching semantics, states that for race-free executions the versioned semantics is equivalent to the shared-memory semantics.

\begin{definition}[Same expressions]
  \[
    \text{same-exprs}(\Conf{T_{sh}}{\_}{\_}, \Conf{T_v}{\_}{\_}) \triangleq
    \forall t. \left(T_{sh}(t) = (\mathit{ts}, e) \iff \exists \delta, c.~ T_v(t) = (\mathit{ts}, e, \delta, c)\right)
  \]
\end{definition}
\begin{theorem}[Equivalence of versioned and shared memory under race-free execution]\label{thm:equiv}
  For a well-formed log that describes both a shared-memory configuration $\mathit{cfg}_{sh}$ and a versioned configuration $\mathit{cfg}_v$, and where each thread in the two configurations have the same expressions, the shared-memory semantics can take a step producing a race-free log $\Log'$ if and only if the versioned semantics can take a step producing the same log.
  \[
    \begin{array}{l}
      \vdash \Log~\mathbf{wf} ~\land~
      \Log \vdash \mathit{cfg}_{sh} ~\land~
      \Log \vdash \mathit{cfg}_v ~\land~
      \text{same-exprs}(\mathit{cfg}_{sh}, \mathit{cfg}_v)~\land~
      \vdash \Log'~\mathbf{rf} ~\implies\\
      \quad(\exists \mathit{cfg}_{sh}'. \mathit{cfg}_{sh} ~|~ \Log \step \mathit{cfg}_{sh}' ~|~ \Log') \iff (\exists \mathit{cfg}_v'. \mathit{cfg}_v ~|~ \Log \step \mathit{cfg}_v' ~|~ \Log')
    \end{array}
  \]
\end{theorem}

The most interesting case of the theorem is reading a variable, where we use \cref{lem:read} to ensure that the same value is read.
For the completeness direction we use \cref{lem:cf-commit} and \cref{lem:cf-merge} to ensure that the versioned semantics do not get stuck on a conflict.
Omitted here is the trivial preservation of $\text{same-exprs}$ in both directions of the equivalence.
Together with the preservation of well-formed describing logs (\cref{thm:preservation}), this extends \cref{thm:equiv} to full executions.

\cref{thm:equiv} shows that for race-free executions, programs have exactly the same semantics under version-controlled memory as under sequentially consistent shared memory.
This ensures backwards compatibility with the JMM for race-free executions.
For a racy execution we either intentionally break backwards compatibility by throwing an exception, or we present a behaviour allowed by the JMM (\cf{} \cref{thm:hbc}).

\subsection{Mechanisation, Including Volatile Variables}
\label{sec:mechanisation}

\newcommand{\LOC}[1]{$\approx$\num{#1}}
We have mechanised our formal development in Rocq~\cite{vcmRocq}, using Arthur Chargu\'{e}raud's TLC library~\cite{tlc} with parts of the proofs written using an LLM.
LLM support was especially helpful for exploring variations of the semantics, such as the different semantics for volatile variables discussed in \cref{sec:volatile}.
The code base weighs in at a total of \LOC{16500} lines, with \LOC{2700} lines of definitions and \LOC{13800} lines of proofs and theorem statements.
The \LinearW{} and \BranchingW{} mechanisations contribute differently: \LinearW{} accounts for \LOC{700} lines of definitions and \LOC{4800} lines of proofs, while \BranchingW{} requires \LOC{1000} lines of definitions and \LOC{6500} lines of proofs, reflecting the higher complexity of its conflict-checking mechanism; the remainder covers the shared memory semantics and definitions shared between semantics.
The presentation in \cref{sec:formal-semantics} defines all three semantics using a single rule set, whereas the mechanisation defines them separately, duplicating overlapping parts and proofs.
The \Read, \Init, and \Write{} functions are replaced by their definitions, splitting the \textsc{Read} rule into a local and a global version for the two versioned semantics.
\Cref{lem:read} is likewise split into separate lemmas for local and global reads.

The mechanised semantics also support volatile variables, following the interpretations listed in \cref{tab:syncasshare-java-intuition} and realised in \cref{tab:vcm-operation-rules}.
The semantics are parameterised over a set of volatile variables, and volatile variables are stored in $\Heap$ alongside regular variables.
The branching semantics additionally maintain a map $V$ of the latest commit of each volatile variable, which is updated on write.
Since the shared semantics is already sequentially consistent, it needs no changes.
The two versioned semantics get an extra premise requiring non-volatility in rules \textsc{Read} and \textsc{Write}, and get two additional rules \textsc{ReadVolatile} (which pulls before reading) and \textsc{WriteVolatile} (which pulls and pushes).
In both semantics, these rules can fail on conflict.
Note also that volatile variables are pushed on write, so they are never in a staging area $\delta$ of any thread.

The mechanised semantics is otherwise the same as the paper, modulo technical details:
the push, pull, commit, and merge operations of the branching model are functions rather than relations, with the conflict-freedom requirement lifted to the rules that use them;
in the linear model commit is likewise a function and the pull is inlined into the rules;
the global case of the \Read{} function is a relation; and
finiteness of maps and sets are threaded through the proofs where needed.

The mechanisation also has a proof of progress for the three semantics, assuming progress for the underlying sequential language.
The progress theorems show that the semantics can get stuck only through a deadlock caused by locks or joins, a failing conflict check, a read of an uninitialised variable, or an attempt to release a lock the thread does not hold.
We consider the two last operations benign since an underlying language could prevent them.

\section{Design Choices}
\label{sec:design}

In this section we discuss some of the design choices inherent to \SyncasShare{}.

\subsection{When are Conflicts Detected?}

The model we have presented detects a write-write conflict at the point where incompatible unordered writes are brought together.
An alternative is to delay this detection until the first read of the ``conflicted'' value.
Programs whose racy writes are never observed would then continue.
Furthermore, if a write to a conflicted location occurs before any read witnesses the conflict, we could allow it to proceed without triggering an error.
This may be a poor choice for testing, but a useful stance for running in production.
This requires tracking conflicted locations in views and defining when later reads force the conflict.
It is important to note that the ability to choose when conflicts are detected is a direct result of \SyncasShare{} using isolated views and explicit communication: in traditional shared memory these updates happen with no visibility.

\subsection{What Constitutes a Conflict?}
\label{sec:what-is-a-conflict}

\SyncasShare{} gives us the further ability to explore the design space of what constitutes a conflict.
In this paper, we have chosen a relatively conservative approach where conflict checking is made between two writes and is value-agnostic.
In \LinearW{}, a thread can push a staged write only if it has seen all later versions of every location it modified.
In \BranchingW{}, merging two unordered branches rejects any location changed in both.
Both definitions ignore the values being written, so two unordered writes of the same value to the same location are treated as incompatible when a later synchronisation action brings the views together.

A more conservative approach would consider the values read by a thread when determining conflicts: when thread's views are brought together, conflicts would be detected based on both the values read and written.
This would be closer to the read log that is constructed in transactional memory systems, where conflicts are detected based on the values read and written by transactions.
More programs would be considered conflicting under this stricter definition, potentially highlighting programmer errors.
However, in contrast to write--write conflicts where there are multiple possible values that can be written concurrently, the read value of an isolated thread can only be the last value it observed.
There is no ``race'' in the traditional sense, since a thread's view changes only at its synchronisation actions.
Maintaining a read log for each thread also increases the complexity of the conflict detection mechanism.

We can also relax the conflict definition.
Some deployed Java idioms are intentionally racy but expected to produce correct results.
A standard example is \c{java.lang.String.hashCode()}, which lazily computes the hash on first call and caches it in a private field.
Synchronisation is intentionally omitted from this operation because the hash is derived from immutable state and integer writes are atomic.
All callers compute the same value regardless of ordering, and the field reads and writes cannot be observed to be incoherent.

To address such patterns, a value-sensitive conflict check would let unordered writes to the same location pass when their values are equivalent.
Primitive values are straightforward, objects can either be compared by equality or through a programmer-defined notion of equivalence.
Reference equality admits only writes of the same object, while an object's own \c{equals()} method lets the programmer say which distinct objects count as interchangeable.

Checking by equivalence rather than equality would exclude such ``benign'' conflicts.
The cost is additional non-determinism in whether an execution conflicts, a more complex conflict check, and user code running as part of conflict detection.
Checking by equivalence associates a policy with a type, but an annotation could associate one with a field (\c{volatile} is already such an annotation).

\subsection{What are the Semantics of Java Volatile Fields?}
\label{sec:volatile}

The JMM establishes requirements for \SyncasShare{}'s backwards compatible interpretation of Java's volatile semantics:
A volatile write synchronises-with subsequent \emph{reads} of that field, but not with other writes to it~\cite[\S17.4.4]{JMM};
Two writes to the same variable unordered by happens-before are therefore a data race~\cite[\S17.4.5]{JMM}, even when that variable is volatile;
And, every volatile write that precedes a read in synchronisation order synchronises with that read, not only the write whose value the read sees~\cite[\S17.4.4]{JMM}.
These guarantees are specific to Java.
In C++ release--acquire semantics, for example, an acquire synchronises only with the release that it takes its value from~\cite{Batty2011}, so a design instantiated in C++ could target a weaker constraint.

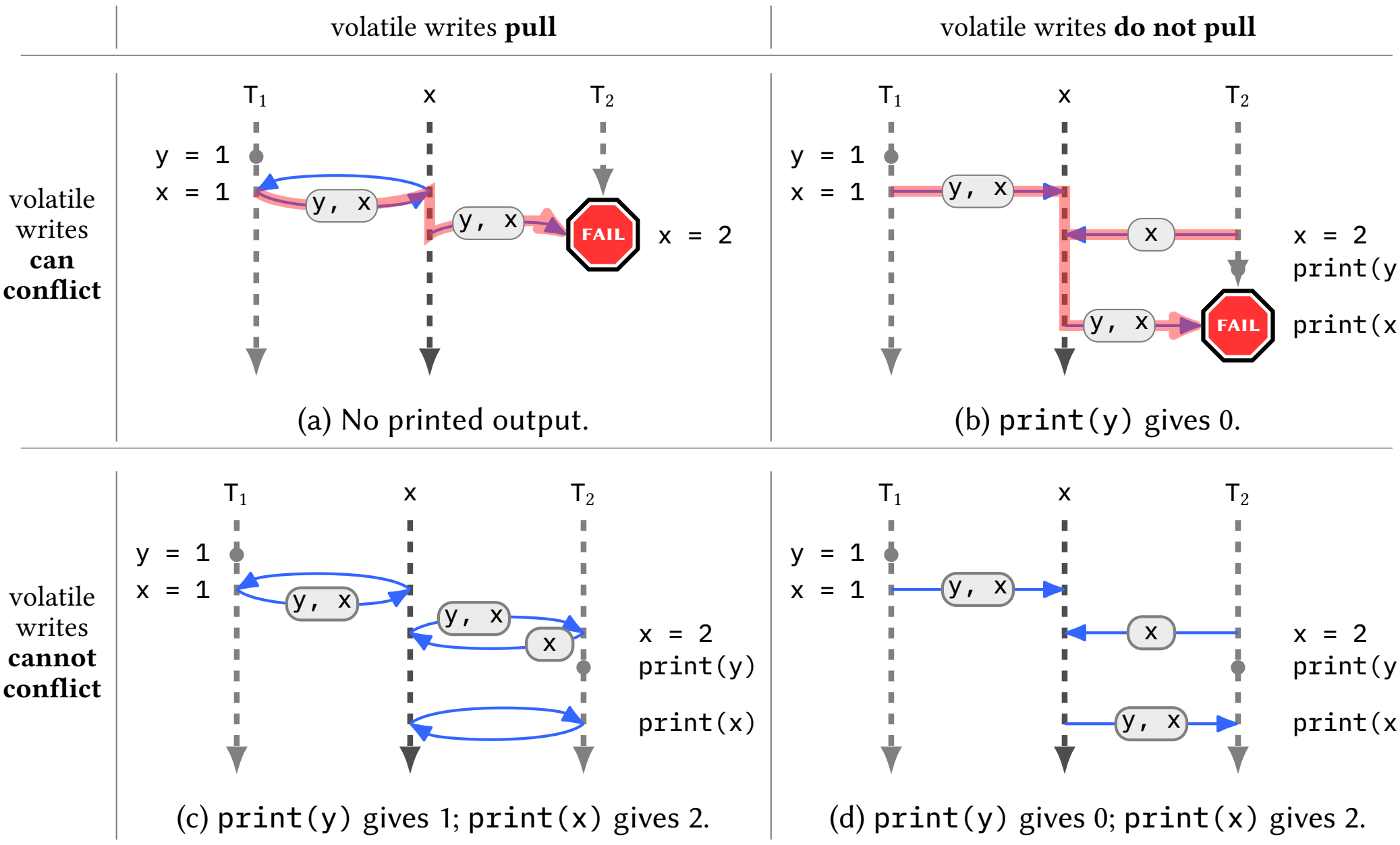
\begin{figure}[t]
  \begin{subfigure}[t]{0.28\linewidth}
    \begin{smallcode}[title={Shared state}, numbers=none, autogobble]
      static volatile int x;
      static int y;
    \end{smallcode}
  \end{subfigure}
  \begin{subfigure}[t]{0.14\linewidth}
    \begin{smallcode}[title={Thread 1}, numbers=none, autogobble]
      y = 1;
      x = 1;
    \end{smallcode}
  \end{subfigure}
  \begin{subfigure}[t]{0.18\linewidth}
    \begin{smallcode}[title={Thread 2}, numbers=none, autogobble]
      x = 2;
      print(y);
      print(x);
    \end{smallcode}
  \end{subfigure}

  \medskip

  \newcommand{\DesignRow}[1]{\raisebox{0.8cm}{\footnotesize\renewcommand{\arraystretch}{0.85}\begin{tabular}[b]{@{}c@{}}#1\end{tabular}}}
  \newcommand{\AxisScope}[1]{{\footnotesize\mdseries #1}}
  \arrayrulecolor{black!50}
  \begin{tabular}{@{}c@{\hspace{0.4em}}|@{\hspace{0.4em}}c@{\hspace{0.4em}}|@{\hspace{0.4em}}c@{}}
    & {\footnotesize \AxisScope{volatile writes} \textbf{pull}}
    & {\footnotesize \AxisScope{volatile writes} \textbf{do not pull}} \\
    \cmidrule(lr){1-3}
    \noalign{\vskip 0.2em}
    \DesignRow{volatile\\writes\\\textbf{can}\\\textbf{conflict}}
  &
  \begin{subfigure}[t]{0.44\textwidth}
    \centering
    \SetLeftCodeShift{\CodeWidthOf{y = 1}}
    \SetRightCodeShift{14pt}
    \begin{tikzpicture}[common timeline styles, y=0.85cm]
      \TimelineColorSetup
      \newcommand{\onex}{0}
      \newcommand{\gx}{1.7}
      \newcommand{\twox}{3.4}
      \node at (\onex,0.6) {T$_1$};
      \node at (\gx,0.6)   {x};
      \node at (\twox,0.6) {T$_2$};
      \DefineThreadEventMacros

      \ThreadEvent{t1}{-0.1}{t1y}{y = 1}
      \ThreadSyncEventFrom{t1}{t1y}{-0.4}{t1x}{x = 1}

      \ThreadSyncEvent{t2}{-1.0}{t2x}{x = 2}
      \ConflictEvent{t2x}{t2conflict}

      \draw[thread lane, ->] (\onex,0.3) -- (\onex,-2.65);
      \draw[shared lane, ->] (\gx,0.3)   -- (\gx,-2.65);
      \draw[thread lane, ->] (\twox,0.3) -- (t2conflict);

      \draw[link oneway] (\gx,|-t1x) to[looseness=.5, out=140, in=30] (t1x);
      \draw[link oneway] (t1x) to[looseness=.5, out=320, in=210]
        coordinate (pushmsg) (\gx,|-t1x);
      \draw[link oneway] (\gx,|-t2x) to[looseness=.5, out=40, in=150]
        coordinate[pos=0.45] (pullmsg) (t2conflict.west);

      \begin{scope}[transparency group, opacity=0.5]
        \draw[conflict] (t1x) to[looseness=.5, out=320, in=210] (\gx,|-t1x)
                              -- (\gx,|-t2x)
                              to[looseness=.5, out=40, in=150] (t2conflict.west);
      \end{scope}
      \node[msg] at (pushmsg) {y, x};
      \node[msg] at (pullmsg) {y, x};

      \ConflictEvent{t2x}{t2conflicttop}
    \end{tikzpicture}
    \caption{No printed output.}
    \label{fig:volatile-design-a}
  \end{subfigure}
  &
  \begin{subfigure}[t]{0.44\textwidth}
    \centering
    \SetLeftCodeShift{\CodeWidthOf{y = 1}}
    \SetRightCodeShift{14pt}
    \begin{tikzpicture}[common timeline styles, y=0.85cm]
      \TimelineColorSetup
      \newcommand{\onex}{0}
      \newcommand{\gx}{1.7}
      \newcommand{\twox}{3.4}
      \node at (\onex,0.6) {T$_1$};
      \node at (\gx,0.6)   {x};
      \node at (\twox,0.6) {T$_2$};
      \DefineThreadEventMacros

      \ThreadEvent{t1}{-0.1}{t1y}{y = 1}
      \ThreadSyncEventFrom{t1}{t1y}{-0.4}{t1x}{x = 1}

      \ThreadSyncEvent{t2}{-1.0}{t2x}{x = 2}
      \ThreadEventFrom{t2}{t2x}{-0.4}{t2py}{print(y)}
      \ThreadSyncEventFrom{t2}{t2py}{-0.65}{t2px}{print(x)}
      \ConflictEvent{t2px}{t2conflict}

      \draw[thread lane, ->] (\onex,0.3) -- (\onex,-2.65);
      \draw[shared lane, ->] (\gx,0.3)   -- (\gx,-2.65);
      \draw[thread lane, ->] (\twox,0.3) -- (t2conflict);

      \draw[link oneway] (t1x) -- (\gx,|-t1x);
      \draw[link oneway] (t2x) -- (\gx,|-t2x);
      \draw[link oneway] (\gx,|-t2px) -- (t2conflict);

      \begin{scope}[transparency group, opacity=0.5]
        \draw[conflict, -] (\onex,|-t1x) -- (\gx,|-t1x) -- (\gx,|-t2x);
        \draw[conflict, -] (\twox,|-t2x) -- (\gx,|-t2x);
        \draw[conflict] (\gx,|-t2x) -- (\gx,|-t2px) -- (t2conflict);
      \end{scope}
      \node[msg] at ($(t1x)!0.5!(\gx,|-t1x)$) {y, x};
      \node[msg] at ($(t2x)!0.5!(\gx,|-t2x)$) {x};
      \node[msg] at ($(\gx,|-t2px)!0.4!(t2conflict.west)$) {y, x};

      \ConflictEvent{t2px}{t2conflicttop}
    \end{tikzpicture}
    \caption{\c{print(y)} gives 0.}
    \label{fig:volatile-design-b}
  \end{subfigure}
  \\[0.35em]
    \cmidrule(lr){1-3}
    \noalign{\vskip 0.35em}
    \DesignRow{volatile\\writes\\\textbf{cannot}\\\textbf{conflict}}
  &
  \begin{subfigure}[t]{0.44\textwidth}
    \centering
    \SetLeftCodeShift{\CodeWidthOf{y = 1}}
    \SetRightCodeShift{14pt}
    \begin{tikzpicture}[common timeline styles, y=0.85cm]
      \TimelineColorSetup
      \newcommand{\onex}{0}
      \newcommand{\gx}{1.7}
      \newcommand{\twox}{3.4}
      \node at (\onex,0.6) {T$_1$};
      \node at (\gx,0.6)   {x};
      \node at (\twox,0.6) {T$_2$};
      \DefineThreadEventMacros

      \ThreadEvent{t1}{-0.1}{t1y}{y = 1}
      \ThreadSyncEventFrom{t1}{t1y}{-0.4}{t1x}{x = 1}

      \ThreadSyncEvent{t2}{-1.0}{t2x}{x = 2}
      \ThreadEventFrom{t2}{t2x}{-0.4}{t2py}{print(y)}
      \ThreadSyncEventFrom{t2}{t2py}{-0.65}{t2px}{print(x)}

      \draw[thread lane, ->] (\onex,0.3) -- (\onex,-2.65);
      \draw[shared lane, ->] (\gx,0.3)   -- (\gx,-2.65);
      \draw[thread lane, ->] (\twox,0.3) -- (\twox,-2.65);

      \PullPushMsg{(t1x)}{(\gx,|-t1x)}{y, x}
      \draw[link oneway] (\gx,|-t2x) to[looseness=.5, out=40, in=150]
        node[message, pos=0.4] {y, x} (t2x);
      \draw[link oneway] (t2x) to[looseness=.5, out=220, in=330]
        node[message, pos=0.25] {x} (\gx,|-t2x);
      \PullPush{(t2px)}{(\gx,|-t2px)}

    \end{tikzpicture}
    \caption{\c{print(y)} gives 1; \c{print(x)} gives 2.}
    \label{fig:volatile-design-c}
  \end{subfigure}
  &
  \begin{subfigure}[t]{0.44\textwidth}
    \centering
    \SetLeftCodeShift{\CodeWidthOf{y = 1}}
    \SetRightCodeShift{14pt}
    \begin{tikzpicture}[common timeline styles, y=0.85cm]
      \TimelineColorSetup
      \newcommand{\onex}{0}
      \newcommand{\gx}{1.7}
      \newcommand{\twox}{3.4}
      \node at (\onex,0.6) {T$_1$};
      \node at (\gx,0.6)   {x};
      \node at (\twox,0.6) {T$_2$};
      \DefineThreadEventMacros

      \ThreadEvent{t1}{-0.1}{t1y}{y = 1}
      \ThreadSyncEventFrom{t1}{t1y}{-0.4}{t1x}{x = 1}

      \ThreadSyncEvent{t2}{-1.0}{t2x}{x = 2}
      \ThreadEventFrom{t2}{t2x}{-0.4}{t2py}{print(y)}
      \ThreadSyncEventFrom{t2}{t2py}{-0.65}{t2px}{print(x)}

      \draw[thread lane, ->] (\onex,0.3) -- (\onex,-2.65);
      \draw[shared lane, ->] (\gx,0.3)   -- (\gx,-2.65);
      \draw[thread lane, ->] (\twox,0.3) -- (\twox,-2.65);

      \draw[link oneway] (t1x) -- node[message] {y, x} (\gx,|-t1x);
      \draw[link oneway] (t2x) -- node[message] {x} (\gx,|-t2x);
      \draw[link oneway] (\gx,|-t2px) -- node[message] {y, x} (t2px);
    \end{tikzpicture}
    \caption{\c{print(y)} gives 0; \c{print(x)} gives 2.}
    \label{fig:volatile-design-d}
  \end{subfigure}
  \\
  \end{tabular}
  \arrayrulecolor{black}
  \caption{One program under the different design choices for volatile fields.}
  \Description[One program under four designs for volatile fields.]{Thread 1 writes the ordinary field y and then the volatile x. Thread 2 writes the volatile x, then prints y and x. In the top row writes to x conflict, and the pull decides whether the conflict is raised at thread 2's write or at its later read. In the bottom row they do not, both executions run to completion, and the pull decides whether thread 2 prints 1 or 0 for y.}
  \label{fig:volatile-designs}
\end{figure}

When reinterpreting volatile reads and writes as communication operations, we have a choice in how to realise the guarantees they provide.
Analogous to monitors, a thread writing to a volatile field pushes its updates into the field's view, where later readers pull them in.
The question is whether the writing thread pulls that view in as well.
If the thread pulls, the write to the volatile field brings the two views together and detects any conflict there.
If the thread only pushes, the views will be brought together in the field's view and next read of the volatile field will be where conflicts are detected.
\cref{fig:volatile-design-a,fig:volatile-design-b} show the effects of this choice.

Even though the JMM considers unordered volatile writes racy, whether or not to \emph{report} them as a conflict is a design choice for \SyncasShare{}.
One option is to relax the conflict semantics for the volatile field whose access brings two views together, along the lines of the equivalence policies in \cref{sec:what-is-a-conflict}.
We could relax it further and resolve unordered writes to the field with a last-writer-wins policy.
The \c{volatile} annotation would then gain a further role, marking the fields where unordered writes resolve rather than conflict.
\Cref{fig:volatile-design-c,fig:volatile-design-d} show this, where Thread~2's \c{x = 2} overwrites Thread~1's \c{x = 1} because \c{x} is excluded from conflict checking.
Note that the full views are still communicated.
In \cref{fig:volatile-design-c} Thread~2 sees the non-volatile \c{y}'s value due to its write to \c{x}.
Had there been other conflicts in the views, an exception would still be raised.
Only the written \c{x} is exempt.

To keep things simple for this paper, we adopt the most conservative approach for volatile fields:
Threads pull and push on volatile writes, and volatile fields are included in conflict checking.
This corresponds to \cref{fig:volatile-design-a}.
A correctly synchronised program behaves the same under all four designs (mechanically verified in Rocq).
The designs give different outcomes only for programs that are incorrectly synchronised.
We will be able to relax this as we learn more about programming styles, the complexity of conflict checking, and the constraints imposed by the memory model.

\subsection{Performance Costs and Opportunities}

\SyncasShare{} will incur runtime costs in both memory and performance.
Conflict checking, maintaining private views, tracking version histories, and communicating between views at synchronisation points all carry a price, and we have not measured it.
However, \SyncasShare{} can remove existing costs as well.
For example, JVMs today guard object construction with fences, so that racing reads cannot observe partially initialised objects.
And compiler optimisations, program transformations, and garbage collection must reason about what other threads can observe, which forces conservative choices and more expensive analyses.
Under \SyncasShare{} a thread sees another's writes only when a synchronisation action communicates them, so today's construction fence would have nothing to guard against.
And as thread execution operates on a private view, a thread's code between synchronisation points is sequential, which simplifies many of these optimisations and enables others.
Evaluating these benefits and reducing what \SyncasShare{} pays for its safety is the main direction for future work.

\section{Related Work}
\label{sec:related-work}

\paragraph{Memory models and DRF guarantees.}
The memory models of Java and C/C++ define concurrency through access-level executions, synchronisation order, visibility, and data-race freedom~\cite{JMM,BoehmAdve2008,Batty2011}.
Earlier Java work showed the difficulty: the original model admitted behaviours that were too weak for programmers and too awkward for compiler writers, leading to the revised JMM and its causality test cases~\cite{Pugh2000,JMMCausalityTests}.
Research continues to test the implementations that aim to realise the JMM~\cite{jMT}.
Hardware and language memory-model work has refined this access-based style by specifying consistency guarantees, compiler transformations, and out-of-thin-air restrictions~\cite{AdveGharachorloo1996,Vafeiadis2015C11Optimisations,PromisingSemantics,SevcikAspinall2008}.
\SyncasShare{} preserves the behaviours of programs that are correctly synchronised under the JMM but changes the treatment of incorrect synchronisation.
A conflict throws an exception, and the JMM has no such behaviour.
An axiomatic presentation of \SyncasShare{} would let us compare the two models directly, and determine whether the JMM already allows everything else \SyncasShare{} does.

\paragraph{Race-exception semantics.}
Goldilocks throws a \c{DataRaceException} from within a Java runtime, backed by precise lockset-based detection~\cite{Goldilocks}.
DRFx and Conflict Exceptions instead raise exceptions on conflicts between concurrently executing regions delimited by synchronisation operations, using hardware support, and Valor achieves the same region-conflict guarantee in software~\cite{DRFx,ConflictExceptions,Valor}.
Like \SyncasShare{}, DRFx detects dynamically and delimits its regions at synchronisation actions.
DRFx compares the individual accesses of two regions and faults when the regions overlap in time, so which pair of conflicting writes it reports depends on the schedule and on how much parallelism the machine provides.
Under \SyncasShare{} a conflict arises where two private updates to the same variable meet at a synchronisation action, which the program's synchronisation structure determines rather than the configuration it runs on.
DRFx monitors a shared-memory execution and guarantees that every execution either faults or is sequentially consistent.
\SyncasShare{} changes what the execution is, and \cref{thm:equiv} guarantees that conflict-free executions coincide with shared-memory behaviour.

\paragraph{Dynamic race detection.}
Dynamic race detectors report executions with conflicting accesses that are unordered by a chosen race relation, in the lockset style established by Eraser~\cite{eraser} or the happens-before style of FastTrack~\cite{FastTrack}, with sampling and JVM-integrated systems such as MaTSa reducing overhead further~\cite{LiteRace, 10.1145/3759548.3763369}.
Predictive detectors and model checkers instead search for races the observed schedule did not exhibit, through weak-causally-precedes and later prediction techniques, Java PathFinder, and optimal dynamic partial-order reduction~\cite{10.1145/3062341.3062374,10.1145/3385412.3385993,10.1145/3597503.3639099,havelund2000model,10.1145/3498711}.
These tools report a race, but once a program is incorrectly synchronised the JMM no longer requires its behaviours to be sequentially consistent, so compiler, JIT, and hardware transformations may expose stale values, reordered effects, or cached reads~\cite{JMM}.
The programmer may therefore still need to reconstruct the failure through the same low-level semantics that made the bug difficult to understand.
Under \SyncasShare{} the program's synchronisation determines the outcome, leaving a thread isolated with its own values or raising a conflict where incompatible views meet.


\paragraph{Concurrent revisions and isolated workspaces.}
Concurrent revisions give forked tasks isolated snapshots and merge them at joins, with isolation types controlling deterministic conflict resolution~\cite{burckhardt2011semantics}, and deterministic threading and workspace-consistency systems similarly use isolated workspaces and controlled commit points~\cite{dthreads, workspaceconsistency}.
Revisions communicate at fork and join, where \SyncasShare{} communicates at every synchronisation action a Java program already performs, including monitor operations and volatile accesses, with ordinary accesses staying private in between.
Restricting communication to fork and join earns revisions their determinism, and \SyncasShare{} gives that up, with conflicts \emph{detected and an exception thrown}, value-agnostically, rather than \emph{merged and resolved} through declared isolation types.
We discussed alternatives which allow for different resolutions in \cref{sec:what-is-a-conflict}.
Distributed version-control systems and libraries such as Irmin expose clone, merge, and conflict-resolution operations over persistent data~\cite{irmin}, where our use of version control is semantic.
\TightenPar{1}

\paragraph{Transactional memory.}
Transactional memory makes blocks of code appear atomic by detecting conflicts in read/write sets and using abort, retry, or rollback to find a serialisable execution~\cite{HTM,STM}, and snapshot isolation and STM algorithms such as TL2 and NOrec maintain version information to validate reads and writes~\cite{SI,TL2,NOrec}.
Snapshot isolation over the whole heap would be a means to achieve \LinearW{}, with each synchronisation action opening a transaction, the next action closing it, and a write--write conflict throwing instead of aborting.
\BranchingW{} has no such encoding, since snapshot isolation commits to a single version history where a synchronisation action communicates through the synchronisation object it names.

\paragraph{Commutativity and transactional meta-theories.}
A line of transactional work makes conflict detection less conservative by using the commutativity of abstract operations rather than read/write sets, reducing false conflicts and the rollbacks they cause~\cite{TransactionalBoosting,CoarseGrainedTransactions}.
The push/pull model of transactions is a meta-theory of serialisability in which a transaction pushes its effects into a shared log and pulls in those of others, with side conditions resolved through commutativity~\cite{PushPull}.
\SyncasShare{} shares this push/pull structure of private and shared views, but reinterprets existing code rather than transforming it, and has no atomicity, serialisability, or rollback.
\SyncasShare{} compares writes to the same variable, which is more conservative.

\paragraph{Type systems for race freedom.}
A language can rule out unsafe sharing statically, with regions and effects in Deterministic Parallel Java~\cite{DPJ}, ownership in Swift 6, Kappa and SafeJava~\cite{swift6,Kappa,PRFJ}, affine types in Rust~\cite{Rust,RustBelt}, and reference capabilities in Pony and behaviour-oriented concurrency~\cite{Pony,BoC,Reggio}.
Lungfish moves some of this enforcement to run time~\cite{Lungfish}, and LVars achieve determinism through monotone writes and lattice joins~\cite{LVars}.
These systems provide static or deterministic guarantees but require programs to be written in, or migrated to, a restricted discipline.
\SyncasShare{} targets the synchronisation constructs already present in existing shared-memory code.

\section{Conclusion}
\label{sec:conclusion}
\SyncasShare{} provides a backwards-compatible and understandable path forward towards data-race-free programming.
Under \SyncasShare{} isolation is the default, so programmers no longer need to work out how ordering in time becomes isolation across space.
Instead, they decide when, and with whom, their threads communicate using existing synchronisation actions.
Further, getting synchronisation wrong no longer corrupts a thread's state, as it simply stays isolated and carries on with its own values.
However, when two private updates to the same globally named variable meet as the result of a synchronisation action, \SyncasShare{} makes the conflict visible.

While implementing \SyncasShare{} will incur overheads, we also see opportunities for savings and new optimisations. We take comfort from the incredible success garbage collection has seen in improving its cost-to-benefit ratio.

\SyncasShare{} is not specific to any language.
Its combination of isolation and explicit communication allows us to expose the key design choices in what is communicated and what constitutes a conflict, which can be different for different languages.
We have explored the trade-offs in two different communication topologies: \LinearW{}, which brings together updates but can leak them between threads that do not directly synchronise, and \BranchingW{}, which keeps updates apart, but at a higher cost in checking.

For concreteness we have instantiated \SyncasShare{} for Java and shown how an existing language can achieve data-race freedom.
Correctly synchronised Java programs can adopt \SyncasShare{} and keep both their code and their behaviour (\cf{} \cref{thm:equiv}).
Incorrectly synchronised programs can adopt it just as readily, and are told if and where a conflict arises.
More than that, \SyncasShare{} is a roadmap for reinterpreting synchronisation in shared-memory languages to rule out data races by construction.

\bibliographystyle{ACM-Reference-Format}
\bibliography{main}

\end{document}